\documentclass{aa} 
\usepackage{amsmath} 
\usepackage{txfonts}
\usepackage{graphicx} 
\usepackage{natbib} 

\begin{document}


\title{Impacts of stellar evolution and dynamics on the habitable zone: \\ The role of rotation and magnetic activity}



\author{F. Gallet\inst{1} \and C. Charbonnel\inst{1,2}  \and  L. Amard\inst{1,4} \and S. Brun\inst{3}  \and  A. Palacios\inst{4} \and S. Mathis\inst{3}  }

\offprints{F. Gallet,\\ email: Florian.gallet@unige.ch}

\institute{$^1$ Department of Astronomy, University of Geneva, Chemin des Maillettes 51, 1290 Versoix, Switzerland \\
  $^2$ IRAP, UMR 5277, CNRS and Universit\'e de Toulouse, 14, av. E. Belin, F-31400 Toulouse, France \\
  $^3$  Laboratoire AIM Paris-Saclay, CEA/DRF-Universit\'e Paris Diderot-CNRS, IRFU/SAp Centre de Saclay, F-91191 Gif-sur-Yvette, France \\
  $^4$  LUPM, UMR 5299, Universit\'e Montpellier/ CNRS, Montpellier, France
  }

\date{ Received / Accepted}


\authorrunning{} \titlerunning{}

\abstract{{With the ever growing number of detected and confirmed exoplanets,} the probability to find a planet that looks like the Earth increases continuously.
While it is clear that being in the habitable zone does not imply being habitable, a systematic study of the evolution of the habitable zone is required to account for its dependence upon stellar parameters.} %
{In this article, we aim to provide the community with the dependence of the habitable zone upon the stellar mass, metallicity, rotation, and for various prescriptions of the limits of the habitable zone.} %
{We use stellar evolution models computed with the code STAREVOL, that includes the most {up to date} physical mechanisms of internal transport of angular momentum, and external wind braking, to study the {evolution of the} habitable zone and of the continuously habitable zone limits.} %
{{Mass and metallicity are the stellar parameters that have the most dramatic effects on the habitable zone limits. Conversely, for a given stellar mass and metallicity, stellar rotation has only a marginal effect on these limits {and does not modify the width of the habitable zone}. {Moreover, and as expected on the main-sequence phase and for a given stellar mass and metallicity, the habitable zone limits remain almost constant which confirm the usual assumptions of a relative constancy of these limits during that phase.} The evolution of the habitable zone limits is also correlated to the evolution of the stellar activity (through the Rossby number) that depends on the stellar mass considered. While the magnetic activity has negligible consequence in the case of more massive stars, these effects may have a strong impact on the habitability of a planet around M dwarf stars. Thus, stellar activity cannot be neglected and may have strong impacts on the development of life during the early stage of the continuously habitable zone phase of low-mass stars.} {Using observed trends of stellar magnetic field strength we also constrain the planetary magnetic field (at the zero order) required for a sufficient magnetospheric protection during the whole stellar evolution.} } %
{We explicit for {the first time} the systematic dependence of planet habitability on stellar parameters along the full evolution of low- and intermediate-mass stars. These results can be used as physical inputs for a first order estimation of exoplanetary habitability.} %


\maketitle

\section{Introduction}


With more than {3000} exoplanets detected inside {a} broad range of configurations in terms of distance from the star, size, mass, and atmospheric conditions, the probability to find {more} habitable planets may dramatically increase in near future. 

{Thanks to} the improving accuracy of modern {observational} techniques, such as the radial velocity and transit methods, newly detected planets have continuously reached smaller radius and mass since the first confirmed exoplanet 51 Peg b ($\approx 150~M_{\oplus}$) discovered by \citet{MQ95}. At the beginning, observed exoplanets were only found close to their host star (later named hot Jupiters), but now detection of telluric planets starts to be frequent\footnote{http://exoplanets.org/} e.g. Gliese 581 c \citep{Udry07}, Gliese 581 e \citep{Mayor09}, Kepler-9 d \citep{Holman10}, Kepler-10 b \citep{Batalha11}, Kepler-11 f \citep{Lissauer11}, Kepler-37 d \citep{Borucki11}, alpha Cen B b \citep{Dumusque12}, Kepler-186 f \citep{Quintana14}, Kepler-438 b \citep{Torres15}.

Among these observed objects, there is Kepler-438 b \citep{Torres15} that have been recently confirmed to be inside the habitable zone of its host's star Kepler-438, an M-dwarf star that have a mass of 0.544 $\rm{M}_{\odot}$ and a radius of 0.520 $\rm{R}_{\odot}$. Kepler-438 b is quite close from the Earth's radius (1.12 $R_{\oplus}$), and though to possess a rocky internal structure with a probability of 69.6\% \citep{Torres15}. 
Such peculiar star-planet system is very interesting to analyse since in that configuration, i.e. where the planet is quite close to its star (around 0.16 AU for Kepler-438 b), {both magnetic and} tidal interactions between the two bodies are not negligible \citep[e.g.][]{Stu14,Stu15,Mathis15,BM16} and could imply tidal dissipation, spin-orbit resonance, and synchronisation of the planet rotation period \citep{Heller11,Quintana14}.

Studying and constrain habitability of this kind of star-planet systems will then be quite difficult but interesting as every physical processes that play even a minor role in the close environment of the star will have to be included.

Life can be found in an extremely wide variety of conditions. {Even on Earth, life forms can be found from extremophile up to mesophile organisms}. Since ``life'' is a process rather than a substance, providing a definition for it is {not straightforward}. {{However}, the development of life needs three key ingredients: a source of energy, some organic material, and a solvent \citep{Forget13}.} 
Life form beyond Earth boundary is completely unknown (yet) and naturally leads to habitable zone's conditions described for the kind of organisms found on Earth. However and despite this geocentric definition, the habitable zone is generally defined in astrophysics as the orbital area around a star where liquid water can be present {at the surface of} the planet \citep{Hart76,Kasting93,Leconte15,Porto15}. Indeed, water is one of the main ingredients for organisms to metabolize and reproduce \citep{Kasting10,Gudel14}.

A planet is called habitable when its orbital radius lies inside the habitable zone and the conditions of habitability are {fulfilled}. These conditions are not well defined yet and are linked to complexification processes. For instance, the presence of a planetary magnetosphere is thought to have a strong impact {in helping} complexification through magnetic shielding \citep{Vidotto13,Stu14,Stu15,Vidotto15}. While the {temporal evolution of the} habitable zone is often {not taken into account to study planetary} habitability \citep{Kasting93,Leconte13}, the {earlier} and subsequent evolution of the habitable zone limits closely follow the inherent evolution of the host-star parameters \citep{Porto15}. This evolution can be very fast during peculiar phases of the stellar life, so that a planet seen in the habitable zone at a time $t$ can leave it in few Myrs. This leads to another important concept that is the continuously habitable zone limits which is linked to the time required, for a planet inside the habitable zone, to favour the emergence of complex organisms \citep{Kasting93}. The minimum time required by ``life'' for complexification is not well known and is usually fixed assuming Earth-like life \citep[between 2 and 4 Gyr, see][]{Valle14,Truitt15}. 

Because of these complex habitability conditions, a planet found inside the habitable zone is not necessarily considered as habitable. However, the prime condition to analyse when studying the habitability of a recently detected planet is whether of not this planet is inside the habitable zone. We then definitely need to perfectly comprehend its evolution as a function of {the variation of} the stellar parameters. 

The aim of this article is to study the impact of stellar parameters such as metallicity, mass, and rotation all along the evolution of low- and intermediate mass stars on the limits of the habitable zone \citep[hereafter HZ, see also][]{Porto15}. 
{Such investigation has already been performed by \citet{Valle14}, \cite{Truitt15} and more recently by \cite{Porto15}, but in the case of non-rotating stellar models. Here we investigate for the first time the impact of stellar rotation and look for a correlation between the stellar activity and the evolution of the HZ limits}. 
{We also display more details and extend the results of \cite{Porto15}}. 

The structure of this article is as follows. In Sect. \ref{refgrid} we present the reference grid {of stellar models} that we used to get the evolution of the internal structure and global parameters {that play a role in the current prescriptions for habitability}. In Sect. \ref{HZ} we give the prescriptions we use for the definition of the HZ and its evolution from the early pre-main sequence phase up to the red giant branch as a function of the stellar parameters. {We first study the impact of stellar parameters (including rotation and metallicity ranging from Z=0.0005 to Z=0.0255) along the evolution of a solar-type star in Sect. \ref{hzevol} and explore the range of stellar mass between 0.2 and 2 solar masses in Sect. \ref{massinit}.} We then examine in Sect. \ref{CHZ} the evolution of the continuously habitable zone. Finally, in Sect. \ref{activity} we explore the correlation between stellar activity and evolution of the habitable zone limits and we conclude in Sect. \ref{conc}.

\section{Reference grid of stellar models}

\label{refgrid}

This paper is based on a grid of standard and rotating stellar models  we computed with the code STAREVOL for a range of initial masses between {0.2} and {2}~M$_{\odot}$ and for {four} values of {[Fe/H] = 0.28, 0, -0.49, and -1.42} (corresponding to metallicities Z=0.0255, 0.0134, 4.3$\times 10^{-3}$, and 5$\times 10^{-4}$ respectively).

We refer to \citet{Amard15} for a detailed description of the stellar evolution code STAREVOL and of the physical inputs (equation of state, nuclear reaction, opacities) as well as the mechanisms that impact the internal transport of chemicals and angular momentum. We briefly recall here the main points. 

\subsection{Basic input physics}

{At solar metallicity, the initial helium mass fraction and the mixing length parameters are calibrated on a solar model with the solar mixture of \citet{AsplundGrevesse2009}, and their values are 0.2689 and 1.702 respectively. We kept the same value of the mixing length parameter for all the stellar models.}  Mass loss rate is applied from the first model on the pre-main sequence (hereafter PMS) and estimated using the prescription of \citet{Cranmer11}.

\begin{figure}[ht]
\centering
 \includegraphics[angle= -90,width=0.5\textwidth]{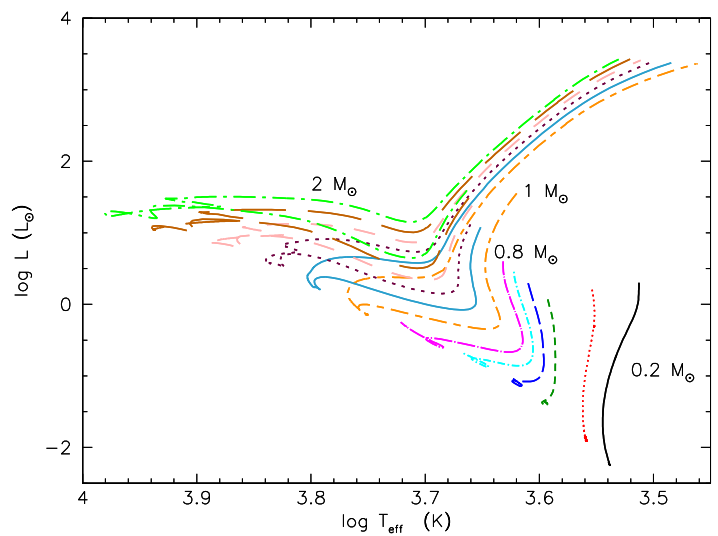}
  \caption{Stellar evolution tracks in the HRD for the standard models of 0.2, 0.3, 0.5, 0.6, 0.7, 0.8, 1.0, 1.2, 1.4, 1.6, 1.8, and 2 $M_{\odot}$ at solar metallicity. {The tracks are shown up to the red-giant branch tip for the more massive stars that have a lifetime lower than 20 Gyr, and up to this age for the lower mass stars.}}
 \label{grid}
\end{figure}

Figure \ref{grid} shows the evolution tracks in the Hertzsprung-Russell diagram (hereafter HRD) of the standard models of masses 0.2, 0.3, 0.5, 0.6, 0.7, 0.8, 1.0, 1.2, 1.4, 1.6, 1.8, and 2 $M_{\odot}$ at solar metallicity.

\subsection{Rotation}

In this work additional rotating models for the 0.3, 1, 1.4, and 2 $M_{\odot}$ stars (see Tab. \ref{vit}) at various initial rotation rate are also considered but not shown in Fig. \ref{grid}.

{The treatment of angular momentum evolution is applied from the first model on the PMS phase using the complete formalism developed by \citet{Zahn92,MaeZah98,MaZa04}, which takes into account advection by meridional circulation and diffusion by shear turbulence \citep[see ][]{Palacios03,Palacios06,Decressin09}. Solid-body rotation is assumed for the convective regions. Magnetic braking is applied from the PMS onward and up to the tip of the red giant branch following \citet{Matt15}.} 

The initial stellar rotation of the solar-like star model is calibrated using the two extreme rotational tracks (fast and slow) from the 2-zones model of \citet{GB15}. For lower and higher masses the observed ratio $r= \Omega_* / \Omega_{\rm{crit}}$ at 1 Myrs, where $\Omega_{\rm{crit}}$ is the critical velocity, is adopted to constrain the initial velocity ($r = 40\%$ and $6\%$ for the fast and slow rotators, respectively).

 In addition to the treatment of transport and loss of angular momentum and mass, the models include the modification of the effective gravity by the centrifugal forces and its effect on the stellar structure equations following \citet{ES76}.

\begin{table}
\caption{Initial velocity used for the stellar rotating tracks at solar metallicity.}
\centering
\label{vit}
\begin{tabular}{ccc}
\hline
\hline
Mass & Initial velocity & Designation\\
($M_{\odot}$) & ($\Omega_{\odot}$) \\
\hline
0.3 & 23 & fast\\
0.3 & 3.8 & slow\\
1.0 & 18.1 & fast\\
1.0 & 3.64 & slow\\
1.4 & 10 & fast\\
1.4 & 2.78 & slow\\
2.0 & 3.97 & fast\\
2.0 & 2.4 & slow\\
\hline   
\end{tabular}
\end{table}

\section{Habitable Zone}
\label{HZ}

\subsection{Definitions}

There are two methods to estimate the location of the HZ limits. The first one relies on the equilibrium temperature calculation that {requires} a strong assumption about the planetary Bond albedo, i.e. the reflectance of the planet's surface (usually taken equal to the Earth's albedo = 0.3). The second method is based on a climate model that directly estimates the albedo as the ratio {between the upward and downward fluxes, inside the planetary atmosphere, that are emitted and received, respectively, by the planet \citep{Kasting93,Kopparapu13,Kopparapu14}}.


The HZ of an exoplanet is generally defined as the area (in terms of distance from its central star) in which a rocky planet, with the right atmosphere, can maintain liquid water at its surface \citep[see ][]{Kopparapu13,Kopparapu14,Linsenmeier15,Torres15}. {Planets complying to this definition are named class I planet by opposition to other type of planets where liquid water can be found under an ice crust \citep[see][]{Lammer09} {with the possibility of a sub-surface habitability}}. The HZ limits (hereafter HZL) are analytically given by \citet{Kasting93} as
\begin{eqnarray}
\label{habdist}
d = \left(  \frac{L/L_{\odot}}{S_{\rm{eff}}} \right)^{0.5} \textrm{AU},
\end{eqnarray}
where $d$ refers to both the inner or outer edge of the HZ (see below), $L$ is the stellar luminosity, $L_{\odot}$ the luminosity of the present Sun, and $S_{\rm{eff}}$ is the effective stellar flux \citep{Kasting93,Kopparapu13,Kopparapu14} {expressed} as the ratio {of} the outgoing IR flux $F_{\rm{IR}}$ {produced by} the planet {to} the net incident stellar flux $F_{\rm{inc}}$ received by the planet {from the star}
\begin{eqnarray}
S_{\rm{eff}} = \frac{F_{\rm{IR}}}{F_{\rm{inc}}}.
\end{eqnarray}
$F_{\rm{IR}}$ is estimated thanks to 1-D radiative-convective climate models such as the one developed by \citet{Kasting93} and \citet{Leconte13}.


Usually, the inner edge of the HZ is chosen to correspond to the runaway greenhouse limit that corresponds to the moment when the surface temperature of the planet is high enough so that the surface of the planet and the base of its atmosphere starts to radiate in both near-IR and visible. These radiations thus increase the outward flux $F_{\rm{IR}}$ while Rayleigh scattering effects, due to {the} vapour content, reduced the absorption of {the} near-IR flux {inside the planetary atmosphere} leading to the saturation of the inward flux $F_{\rm{inc}}$ \citep{Kopparapu13,Kopparapu14}. The surface temperature of the planet then reaches $T_{\rm{surf}} \gg $ 647 K at which water reservoirs are completely evaporated. The legitimacy of this inner limit is presently questioned by the community as at that point the surface temperature of a given planet is way too high to maintain Earth-like life. The runaway greenhouse limit represents a quite optimistic limit and somehow the extreme internal limit of the HZ. The outer edge of the HZ is defined by the maximum greenhouse limit, where the $\rm{CO_2}$ Rayleigh scattering reduces the efficiency of the greenhouse effect. At that limit, the planetary surface temperature is fixed at 273 K \citep[i.e. the freezing point of water,][]{Kasting93,Selsis07,Kopparapu13,Kopparapu14}.

{In this work we refer to the HZL as the location of the inner and outer edge of the HZ around the considered star, while the width of the HZ is the distance between the inner and outer edge of the HZ.}

\subsection{Prescriptions from the literature}

In this section we describe different prescriptions from the literature that are used to estimate the HZL.{We compare in Sect. \ref{prescription} the impact of each of these prescriptions on the HZL.}

\paragraph{{\citet{Kopparapu14}}}
~~\\

\noindent Using the 1-D radiative-convective cloud-free climate model from \citet{Kasting93}, \citet{Kopparapu14} provide an analytic expression for $S_{\rm{eff}}$ {that one can directly use} to compute the HZL. By {using a polynomial fit}, they expressed $S_{\rm{eff}}$ as a function of the effective temperature of the host star $T_{\rm{eff}}$ for different cases of HZL definition (and thus planetary atmosphere) as
\begin{eqnarray}
S_{\rm{eff}} = S_{\rm{eff}\odot} + aT_* + bT_*^2 + cT_*^3 + dT_*^4,
\end{eqnarray}
where $T_* = T_{\rm{eff}}-5780$. $S_{\rm{eff}\odot}$ is {self-consistently} estimated and somehow represents the strength of the greenhouse effect.  $S_{\rm{eff}\odot}$, $a$, $b$, $c$, and $d$ are model parameters that depend on the planetary atmosphere. {Here we use the values listed in Table \ref{Seffin}}.

{For the inner edge, \citet{Kopparapu13} started their modelling with a fully saturated ``Earth'' model and {considered that the flux received by the planet at the top of its atmosphere is the same as that received from the Sun by the Earth} (i.e. 1360 W.m$^{-2}$). Then they increased the surface temperature ($T_s$) of the planet from 200 up to 2200 K and looked at the evolution of the two fluxes, planetary albedo, and $S_{\rm{eff}}$. The moment when $S_{\rm{eff}}$ runaways defines the runaway greenhouse limit.} {For the outer edge, \citet{Kopparapu13} started their simulation with a 1 bar $\rm{N_2}$ planetary atmosphere with a surface temperature of 273 K (i.e. the freezing point of water). Then they increased the $\rm{CO_2}$ partial pressure from 1 to 35 bar and followed the evolution of $S_{\rm{eff}}$ that exhibits a minimum that is reached when the efficiency of the greenhouse effect due to $\rm{CO_2}$ compound is maximum. This minimum is called the maximum greenhouse limit and corresponds to the largest distance, from the star, required to maintain a surface temperature of 273 K on the planet.}

\citet{Kopparapu14} explored three different masses (0.1, 1, and 5 $M_{\oplus}$) and concluded that the larger the planet, the wider the HZL. {In the present work} we only consider the Earth mass planet case.  

\begin{table}[!h]
\caption{Value of the constants given by \citet{Kopparapu14}.}
\centering
\label{Seffin}
\begin{tabular}{ccccccc}
\hline
\hline
& $S_{\rm{eff}\odot}$ &  a ($10^{-4}$) &  b ($10^{-8}$) & c ($10^{-12}$) & d ($10^{-15}$)\\
\hline
inner & 1.107 & $1.332 $ & $1.58$ & $-8.308$ & $-1.931$ \\
outer & 0.356 & $0.6171$ & $0.1698$ & $-3.198$ & $-0.5575$\\
\hline   
\end{tabular}
\end{table}

{All the calculations {reported} in \citet{Kopparapu14} assumed a Sun like star but these prescriptions are applicable, using interpolation, for stars with effective temperatures between 2600 and 7200 K, i.e. between 0.1 and 1.4 $M_{\odot}$ \citep{Kopparapu14}.}

\paragraph{{\citet{Selsis07}}}
~~\\

\noindent \citet{Selsis07} also provide a prescription for the inner and outer edge of the HZ that is based on the various cases of planetary atmosphere explored by \citet{Kasting93}. {They use the following expression for the effective stellar flux :}
\begin{eqnarray}
S_{\rm{eff}} = \frac{1}{\left( S_{\rm{eff}\odot} - aT_* - bT_*^2\right)^2} 
\end{eqnarray}
where $T_* = T_{\rm{eff}} - 5700$. The values of $S_{\rm{eff}\odot}$,  a, and  b are given in Table \ref{Seffin2}. For the inner edge we used $S_{\rm{eff}\odot,in} = 0.84$ in the case of a runaway greenhouse limit (0.95 corresponds to the $T_{\rm{surf}} = 373$ K limit).
\begin{table}[!h]
\caption{Value of the constants given by \citet{Selsis07}.}
\centering
\label{Seffin2}
\begin{tabular}{ccccc}
\hline
\hline
& $S_{\rm{eff}\odot}$ &  a ($10^{-4}$) &  b ($10^{-9}$) \\
\hline
inner & 0.84-0.95  & $0.27619$ & $3.8095$ \\
outer & 1.67 & $1.3786$ & $1.4286$\\
\hline   
\end{tabular}
\end{table}

{The parabolic fitting of \citet{Selsis07} is here scaled to the values found by \citet{Kasting93} in the case of a G2 star and can be extrapolated to stars with $\rm{T_{eff}}$ between 3700 and 7200 K.} 

\paragraph{{\citet{Underwood03}}}
~~\\

\noindent We finally give the HZ prescription from \citet{Underwood03}, {who also use a polynomial fit to express the effective stellar flux $S_{\rm{eff}}$}
\begin{eqnarray}
S_{\rm{eff}} = S_{\rm{eff}\odot} + aT_{\rm{eff}} + bT_{\rm{eff}}^2,
\end{eqnarray}
where $T_{\rm{eff}}$ is the effective stellar surface temperature, and the values of $S_{\rm{eff}\odot}$, $a$, and $b$ are summarized in Table \ref{Seffin3}).

\begin{table}[!h]
\caption{Value of the constants given by \citet{Underwood03}.}
\centering
\label{Seffin3}
\begin{tabular}{ccccc}
\hline
\hline
& $S_{\rm{eff}\odot}$ &  a ($10^{-4}$) &  b ($10^{-8}$) \\
\hline
inner & 1.268  & $-2.139$ & $4.190$ \\
outer & 0.2341 & $-0.1319$ & $0.619$\\
\hline   
\end{tabular}
\end{table}

{Using the different cases provided by \citet{Kasting93}, \citet{Underwood03} estimate the effective stellar flux as a function of the effective temperature between 3700 K (M0 star) and 7200 K (F0 star). }

~~\\
Among the three expressions above, \citet{Selsis07} and \citet{Underwood03} are both based on the work of \citet{Kasting93} and the difference that they may have is a consequence of the fitting method employed to express the stellar flux $S_{\rm{eff}}$ as a function of $T_{\rm{eff}}$. {The main difference between \citet{Kopparapu14} and the other two prescriptions lies in an update of the \citet{Kasting93} model by \citet{Kopparapu14} that includes the most recent version of the HITEMP opacity database including more $\rm{H_2O}$ lines in the near-infrared \citep{Kopparapu13}.} {The impact of these prescriptions are studied in Sect. \ref{prescription}. {After this comparison, we use the \citet{Kopparapu14} prescription}.}

\subsection{Limitations of the current prescriptions for HZ}

The prescriptions described above are very sensitive to the climate model used to estimate the impact of the incident stellar flux. Therefore, they are only valid in the framework of the specific climate model developed by \citet{Kasting93}. Moreover, the atmospheric compositions used by \citet{Kopparapu14} imply that their HZ's prescriptions are only valid in the case of a planetary atmosphere similar to that of the Earth at the present time. This corresponds to the main sequence phase of our Sun, where the HZ and the stellar parameters are not evolving much and where the stellar activity is quite low. However, the Earth's atmospheric composition has drastically changed during the last 4 Gyr. It is thought that this latter reached its current composition between 400 to 600 Myr ago. In this work  we display the entire HZ evolution anyway in order to highlight the fact that stellar models should be used, especially, during the initial phases of stellar evolution during which stellar parameters heavily and clearly evolve. However, the results will have to be taken with caution, as the evolution of the planetary atmosphere must be taken into account as well, and a complete modelling should involve the co-evolution of both the star and the planetary atmosphere \citep{Leconte13b}.

These prescriptions are also dedicated to stars with effective temperature between 2600-3700 and 7200 K \citep{Underwood03,Selsis07,Kopparapu14} corresponding to stellar mass of 0.1 and 1.4 $M_{\odot}$, respectively. \citet{Kasting93} and \citet{Kopparapu14} chose to stop their analysis at this upper mass limit because stars beyond 1.5 $M_{\odot}$ will have a lifetime shorter than 2 Gyr while it is thought that complex life needs more than 1 Gyr to evolve. However, and in order to have a quite large range of initial stellar mass we extrapolated these prescription for higher mass star. The validity of this extrapolation is of course questionable but the following of this article will show that the evolution of the HZ of a 2 $M_{\odot}$ star is consistent (same temporal features) with that of a 1 $M_{\odot}$ star. Moreover, the habitable zone distance $\rm{d}$ (Eq. \ref{habdist}) have a quite weak dependence on $\rm{T_{eff}}$ (through $\rm{S_{eff}}$). For instance, a change of 20\% of $\rm{S_{eff}}$ will only modified $\rm{d}$ by about 10\%.

Finally, all these quantities are given in the case of a cloud free atmosphere. \citet{Selsis07} studied the impact of the presence of cloud (from a cloud free to a planet fully covered by cloud, using the \citet{Kasting93} model) on the inner and outer edge of the HZ (see their Table 2). While the outer edge of the HZ is not impacted by the percentage of cloud, the inner edge will almost be reduced by a factor of two when the covering percentage goes from 0\% to 100\%. 

\section{Evolution of the habitable zone. The case of a {1 $M_{\oplus}$ planet around an} evolving 1 $M_{\odot}$ host star}
\label{hzevol}

In this section, we focus on the case of a 1 $M_{\oplus}$ planet around a 1 $M_{\odot}$ mass star and study the impact of the intrinsic stellar evolution, metallicity, and rotation on the HZL. {We do not {include the effects of planetary mass since they have been shown to be relatively marginal} \citep[at least for planets with masses below 5 $M_{\oplus}$, see][]{Kopparapu13,Kopparapu14}.}

\subsection{Generalities}

\begin{figure}[ht]
\centering
 \includegraphics[angle= -90,width=0.5\textwidth]{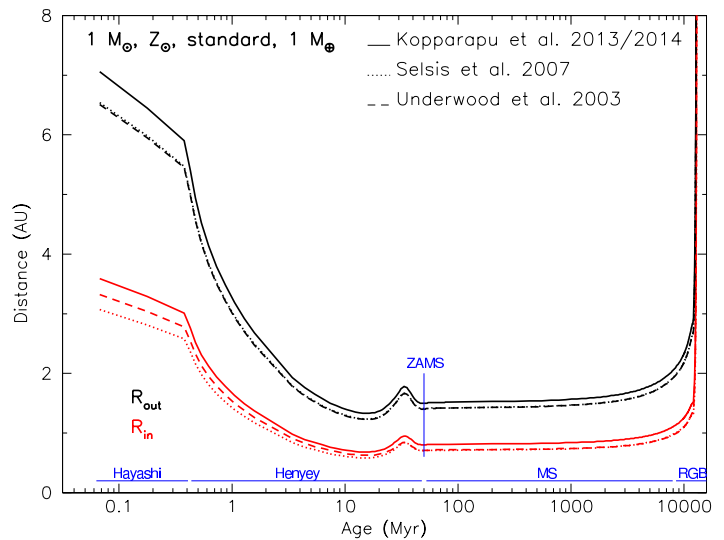}
  \caption{Evolution of the HZL as a function of age for a 1 $M_{\odot}$ star with solar metallicity. The inner and outer edge of the HZ are represented by the red and black lines, respectively. The solid, dotted and dashed lines are associated to HZ prescription from \citet{Kopparapu14}, \citet{Selsis07}, and \citet{Underwood03}, respectively. The blue lines depict the localisation of the different phases : PMS (Hayashi, Henyey), ZAMS, MS, and RGB.}
  \label{HZevol1msol}
\end{figure}

The HZL evolution around a 1 $M_{\odot}$, non rotating, star with solar metallicity {can be followed in Fig. \ref{HZevol1msol} were we show} the temporal evolution of the inner and outer edges of the HZ from the early-pre-main sequence to the tip of the red giant branch. 
This evolution can be separated into several main phases : PMS, Zero-Age Main Sequence (ZAMS), Main Sequence (MS), and post main sequence (e.g. the Red Giant Branch, RGB). 

The PMS phase starts along the so-called Hayashi track \citep{Hayashi61}, where the star evolves at nearly constant effective temperature {and decreasing surface luminosity}. One sees first a rapid and sharp decrease of the size of the HZL that reach a minimum at the end of the Hayashi track {(i.e. between $2\times10^2$ and $1.5\times10^7$ years)}. {Note that} the PMS phase continues along the Henyey track \citep[for stars above 0.5 $M_{\odot}$,][]{Hen55}, where the star evolves at nearly constant luminosity {and increasing effective temperature} towards the ZAMS, defined by the moment during the evolution when the hydrogen burning is dominant in the stellar core i.e. when the hydrogen abundance is lower than about 0.998 of the initial one, that is reached at about 50 Myr for our solar model. During this last PMS stage, the HZL slightly increase and reach a local maximum at ZAMS.

{At the ZAMS the stellar structure eventually reaches a quasi-steady state due to a quasi-hydrostatic equilibrium regime that is ensured by efficient nuclear reactions}. At that point the internal temperature of the star is high enough to initiate complex nuclear reactions, that will then counteract the stellar contraction by balancing the gravity with thermal pressure. During the MS phase (i.e. from {50} Myr to about 10 Gyr) the HZL remain more or less constant at about 1.5 AU for the outer edge and 0.8 AU for the inner edge \citep{Porto15}. {This confirms the general assumptions of a relative constancy of these limits during the MS phase.} 


{Finally, during the RGB phase (i.e. from about 10 Gyr) the star evolves toward higher luminosity and lower effective temperature onward. Correspondingly, the HZL {are shifted} outward reaching values as high as 8 AU to 12 AU. During this final stage, the HZL move continuously away from the star which luminosity increases.}

{The respective durations of each evolutionary phases are clearly unbalanced and will have different kind of effect on the evolution of life itself. For instance, while a 1 $M_{\odot}$ star will pass only about 50 Myr on the PMS phase, it will pass several Gyr on the MS phase. Hence, the PMS phase is clearly too short to be considered as relevant for the development of life. However, to constrain the planetary habitability of a given planet, this very early phase plays a key role through the MHD and tidal star-planet interaction \citep[see][and Sect. \ref{activity} of this article]{Zahn89,Vidotto13,Stu14,Stu15,Mathis15,Mathis152}. Furthermore, as the HZL sharply evolve during the PMS phase, the surface conditions of a planet could also strongly evolve and thus latter constrain the framework in which life will evolve.}

\subsection{Impact of the HZ prescription}
\label{prescription}

In Fig. \ref{HZevol1msol} we also compare the HZL prescriptions from \citet{Underwood03}, \citet{Selsis07}, and \citet{Kopparapu13,Kopparapu14}. {Since all these prescriptions have a common base}, the shape of the HZL evolution is the same for the three prescriptions. \citet{Kopparapu14} produces higher values for the inner (hereafter $R_{\rm{in}}$) and outer (hereafter $R_{\rm{out}}$) edge of the HZ compared to \citet{Selsis07} and \citet{Underwood03}.  While there is no significant differences between \citet{Selsis07} and \citet{Underwood03} for $R_{\rm{out}}$, there is a small shift toward lower values for $R_{\rm{in}}$ from the PMS to the MS phase for the \citet{Selsis07} prescription compared to \citet{Underwood03}.

The prescriptions of \citet{Selsis07} and \citet{Underwood03} are both based on the HZ estimations given in \citet{Kasting93} for three stellar effective temperatures \citep[3700, 5700, and 7200 K, see Table 3 from][]{Kasting93}. The difference of fitting method between the three HZ prescriptions leads, for the inner edge calculated with \citet{Underwood03}, to an higher $R_{\rm{in}}$ at low temperature and to a lower $R_{\rm{in}}$ at high temperature compared to \citet{Selsis07}. Conversely, for the outer edge calculated with \citet{Underwood03}, this leads to a lower $R_{\rm{out}}$ at low temperature and to an higher $R_{\rm{out}}$ at high temperature compared to \citet{Selsis07} (see Fig. \ref{HZevol1msol}).

{{Therefore,} as shown by Fig. \ref{HZevol1msol} the impact of the fitting method on the HZ is non negligible, and even if these prescriptions have a common base \citep[i.e. ][]{Kasting93} these methods lead to slightly different, of about 10\%, estimates of the HZL.}

\subsection{Metallicity effect}
\label{metal}

Metallicity is one of the main {ingredients} that strongly change {the structure and} evolution of a star of a given initial mass.

\begin{figure}[ht]
\centering
 \includegraphics[angle= -90,width=0.5\textwidth]{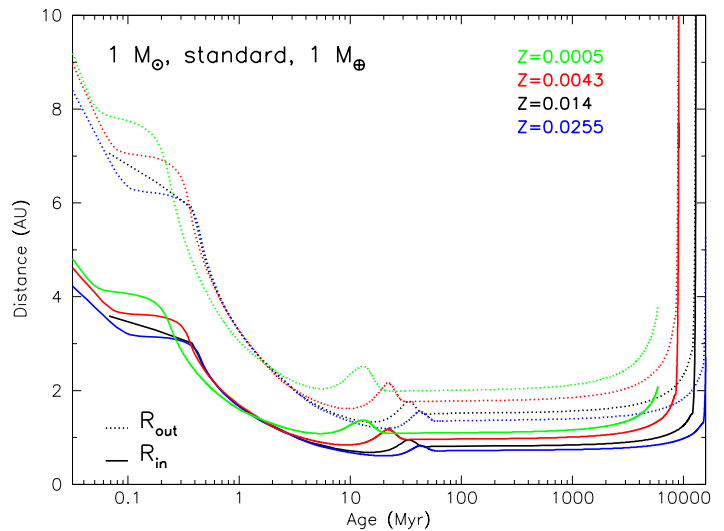}
  \caption{Evolution of the HZ as a function of the stellar age for a 1 $M_{\odot}$ star and for metallicities Z=0.0255, 0.014, 0.0043, and 0.0005 (blue, black, red, and green, respectively). The inner and outer edge of the HZ are represented by the solid and dashed lines, respectively.}
  \label{HZz}
\end{figure}

The major effect of metallicity is to produce, due to opacity effects, a translation in both luminosity and effective temperature along the evolution tracks in the HRD. When the metallicity decreases, the quantity of metallic bound-free absorption inside the star is reduced. This leads to the decrease of the stellar opacity allowing an easier escape of energy and thus increasing the stellar luminosity and temperature. For a given initial stellar mass and at a given evolutionary phase, $T_{\rm{eff}}$ and $L_*$ both increase for decreasing metallicity. This directly produced an increase of the HZL that will almost evolve as $L_*^{0.5}$ \citep[since the HZL do not depend much on the effective temperature according to][]{Kopparapu13,Kopparapu14}. Figure \ref{HZz} shows the evolution of the HZL in the case of a 1 $M_{\odot}$ star for $\rm{[Fe/H]}$ ranging between +0.28 dex (Z=0.0255) and -1.42 dex (Z=0.0005). 

In addition to a shift in luminosity and effective temperature, changing the stellar metallicity also modifies the time at which the characteristic features on the tracks appear. Indeed, the overall lifetime  of a star decreases for decreasing metallicity.  

As shown in \citet{Porto15}, the impact of metallicity is clearly visible at ZAMS. First, the location of the ``bump'' (see Sect. \ref{massinit}) is shifted towards younger ages for decreasing metallicities (see above).  As a result of faster evolution the maximum amplitude of the ``bump'' is reached earlier for decreasing metallicity. Moreover, a decrease in metallicity will move outward the inner and outer edge of the HZ.

\subsection{Impact of stellar rotation}

A star is naturally a rotating body. From its formation inside a molecular cloud up to its nuclear death, the rotation always plays a central role \citep{Maeder09}. {Stellar rotation modifies the stellar effective temperature and luminosity, which are important parameters for the location of the HZ. It also has an indirect impact on the HZ through stellar activity (see Sect. \ref{activity}).} While the evolution of stellar angular velocity is now fairly well known \citep{RM12,GB15,Johnstone15,Amard15} the early-PMS phase and its Star-Disk Interaction (hereafter, SDI) remain quite a mystery \citep{Ferreira13}. Yet, the planetary formation is thought to happen during the first 10 Myrs \citep{Bell13} of the disk's lifetime which may be interrupted by the SDI and then have severe consequences on the first stage of planetary evolution.

\begin{figure}[ht]
\centering
 \includegraphics[angle= -90,width=0.5\textwidth]{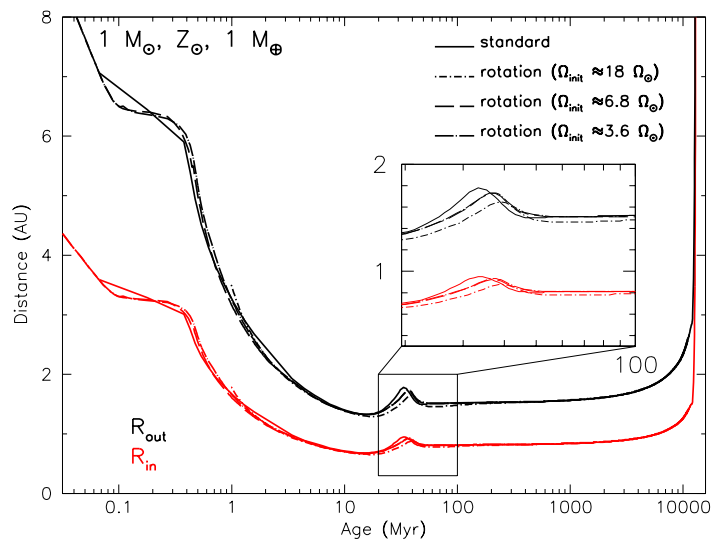}
  \caption{Effect of the change of initial angular velocity on the evolution of the HZ limits in the case of a 1 $M_{\odot}$ star with solar metallicity. Inner and outer edge of the habitable zone are represented by red and black lines, respectively. Standard model is plotted in dotted line and rotation models in dash-dotted, long-dashed, and long-dash-dotted lines for 18, 6.8, and 3.6 $\Omega_{\odot}$ {initial rotation rate}, respectively.}
  \label{HZrot}
\end{figure}

A given star will feel its effective surface temperature and luminosity slightly decreased with increasing rotation rate (there are also marginal effects on the star lifetime that will be slightly reduced with increasing stellar rotation). Figure \ref{HZrot} shows the evolution of the corresponding HZL for a standard model and rotating ones. This figure shows that the effect of a change in rotation rate on the HZL is marginal (the four lines are almost superimposed). {The maximum effect of rotation is seen at ZAMS. Indeed, as the initial rotation rate of given rotating model increases, the corresponding star will reach an higher angular velocity at the end of the contraction phase where the rotation rate reaches a maximal value \citep[see][]{GB13,GB15,Amard15}. As centrifugal effects are directly proportional to the squared of the stellar angular velocity, their impact will also be maximum during this phase.} {This effect decreases as the initial rotation rate decreases since the star tends to reach the standard model case when reducing its rotation rate.}

\subsection{Impact of the micro-physics used in stellar models}

{\citet{Valle14} and \citet{Truitt15} also analyse the impact of stellar parameters on the evolution of the HZL. While \citet{Valle14} based their analysis on the FRANEC code \citep{FRANEC}, \citet{Truitt15} used the TYCHO code \citep{Young05} to calculate the evolution of both luminosity and effective temperature. We compare in this section the impact of the corresponding choice of input micro-physics on the evolution of the HZL.} 

\begin{figure}[ht]
\centering
 \includegraphics[angle= -90,width=0.5\textwidth]{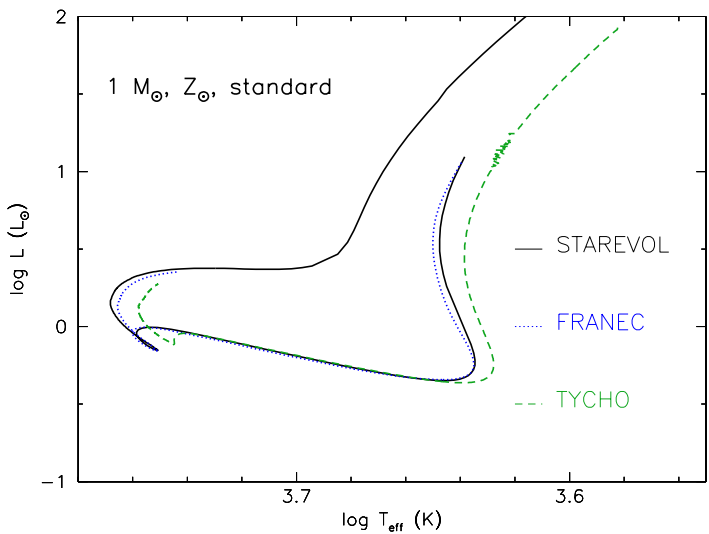}
  \caption{Stellar evolution tracks in the HRD for the standard models of a 1 $M_{\odot}$ at solar metallicity. The black solid, blue dotted and dark green dashed lines are associated to stellar model from this work, \citet{Valle14}, and \citet{Truitt15}, respectively.}
  \label{HRDmodel1msol}
\end{figure}

Figure \ref{HRDmodel1msol} shows the Hertzsprung-Russel diagram of a 1 $M_{\odot}$ star with solar metallicity from the STAREVOL, FRANEC, and TYCHO stellar evolution codes, respectively. The micro-physics used in these last two stellar codes are briefly recalled here. 

\paragraph{{TYCHO : \citet{Young05}}}
~~\\

TYCHO\footnote{TYCHO Interpolation Tool, nicknamed Uranienborg \url{http://bahamut.sese.asu.edu/~payoung/AST_522/Evolutionary_Tracks_Database.html}} and STAREVOL contain quite different micro-physics in terms of chemical mixture, opacities, and nuclear reactions. However, and with the available models, it is not possible to precisely isolate the impact of each of these changes.  For this model the chemical mixture of \citet{L10} was used with a metallicity $Z = 0.0153$ and an initial helium mass fraction $Y = 0.2735$.

\paragraph{{FRANEC : \citet{Tognelli2011}}}
~~\\

This latest version of the FRANEC\footnote{FRANEC \url{http://astro.df.unipi.it/stellar-models/index.php?m=8}} code has been upgraded and now includes new abundances and atmospheric conditions. Even if different mixing length parameters are used between STAREVOL and FRANEC (1.702, and 1.68, respectively), the results of these models are quite close to each other as shown by Fig. \ref{HRDmodel1msol}. For the FRANEC model, the chemical mixture of \citet{Asplund05} was used with a metallicity $Z = 0.01377$ and an initial helium mass fraction $Y = 0.2533$.

\begin{figure}[ht]
\centering
 \includegraphics[angle= -90,width=0.5\textwidth]{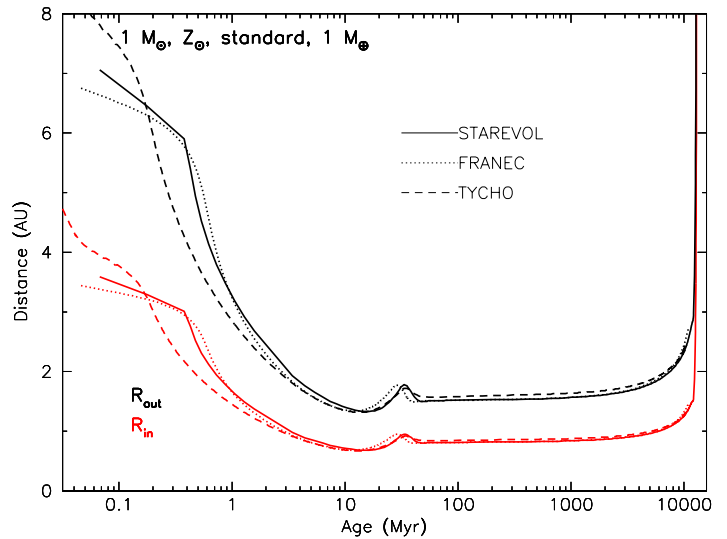}
  \caption{Evolution of the HZL as a function of age for a 1 $M_{\odot}$ with solar metallicity. The inner and outer edge of the HZ are represented by the red and black lines, respectively. The solid, dotted and dashed lines are associated to stellar models from this work, \citet{Valle14}, and \citet{Truitt15}, respectively.}
  \label{HZmodel1msol}
\end{figure}

As shown by Fig. \ref{HRDmodel1msol} the quite large difference of metallicity between TYCHO and STAREVOL/FRANEC results in an cooler stars (at a given luminosity) for TYCHO compared to STAREVOL/FRANEC. As the dependence of the HZL on the effective temperature is small \citep[see][]{Kopparapu13,Kopparapu14} the evolution of the HZL related to these three stellar codes is expected to be quite similar (see Fig. \ref{HZmodel1msol}.) 

\subsection{Summary}

{In order to analyse the impact of stellar parameters and stellar models on the evolution of the HZ we focused on an Earth-mass planet orbiting a solar-type star. To have the best estimation of the location of the HZ we first reviewed the impact of numerical {modelling} considerations, such as the choice of stellar model, and then evaluated the impact of a change in the stellar physical parameters on the evolution of the inner and outer edge of the HZ.}

As we shown, a change of HZ prescription will have a small impact on the HZL. The HZ prescription of \citet{Selsis07} and \citet{Underwood03} will be on average around 10\% closer to the star compared to the HZ prescription of \citet{Kopparapu14}.

{From the point of view of the physical stellar properties, we showed that a change, even small, in metallicity dramatically impacts the location of the HZ, especially beyond the ZAMS and during the entire MS phase. A decrease in metallicity moves the HZ outward (because $L_*$ and $T_{\rm{eff}}$ both increase) and shortens the stellar evolutionary phases. With a change of metallicity of almost 100\% (Z=0.0255 to Z=0.0005) the inner and outer edge moved outward, at ZAMS, by about 50\%. Rotation marginally affects the HZ and its effects are the strongest when the stellar rotation is maximum i.e. at the end of the contraction phase. During this phase, the effect of rotation is only to move the HZ closer to the star of about 10\% and will only last few tenth of Myrs in the case of a solar-type star.}

{The impact of micro-physics input is not null but quite marginal and only affects the earliest phase of stellar evolution, namely the PMS phase (see Fig. \ref{HZmodel1msol}). {Indeed,} during the most important phase, in terms of duration, (i.e. the MS) the input micro-physics have only minor impact as long as the luminosity during that phase is not much affected.} The TYCHO and FRANEC stellar evolution code produced at ZAMS an HZ that is about 3 and 1\% closer to the star compared to the STAREVOL code. However, a more detailed analysis should be performed so as to extract the specific impact of each of these input micro-physics. 

\section{HZ of a 1 $M_{\oplus}$ planet around stars of various initial masses}
\label{massinit}

In this section we {extend the previous} study {to evaluate} the impact of the stellar mass on the HZL {along stellar temporal evolution.} 

Figure \ref{grid} shows that, at a given evolution phase, luminosity and effective temperature increase for increasing mass. The direct effect {of a mass increase} on the HZL is to shift both the inner and outer edges of the HZ further away from the star as well as to increase the {width} of the HZ.

\begin{figure}[ht]
\centering
 \includegraphics[angle= -90,width=0.5\textwidth]{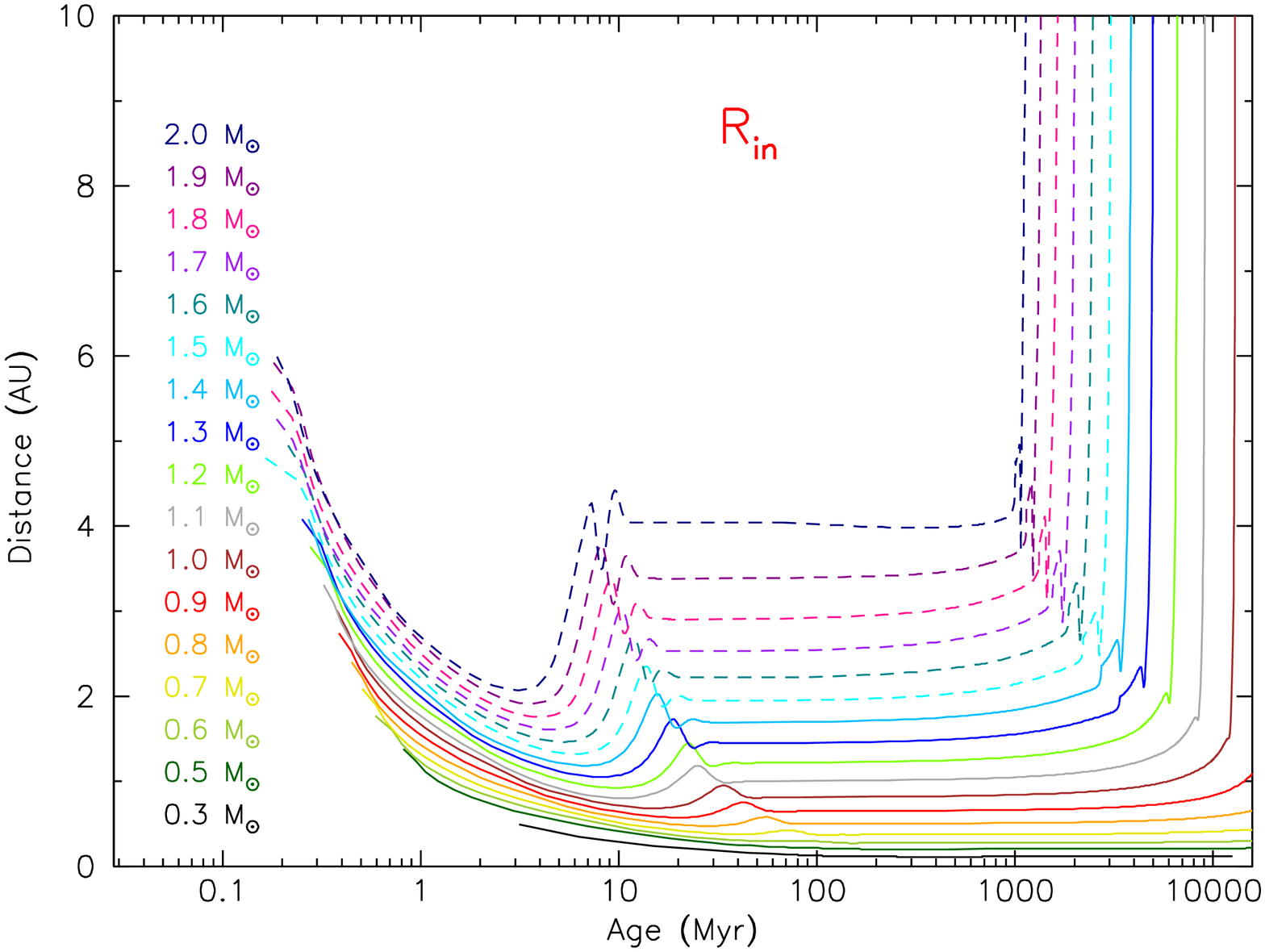} \\
\includegraphics[angle= -90,width=0.5\textwidth]{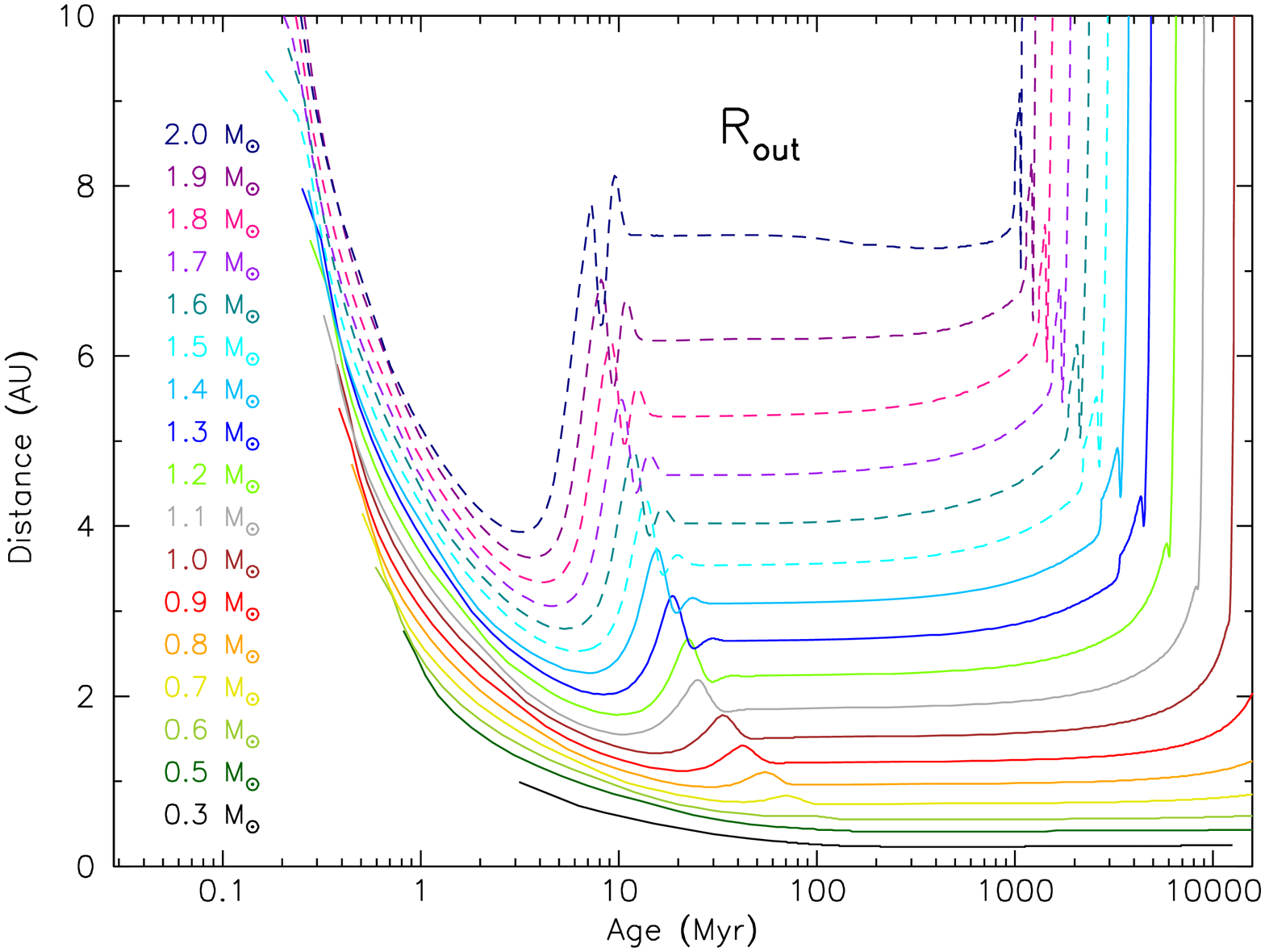} 
\caption{Evolution of the inner edge and outer edges of the HZ (top and bottom respectively) as a function of the time for stars of 0.3 (black), 0.5 (dark green), 0.6 (yellow-green), 0.7 (gold), 0.8 (orange), 0.9 (red), 1.0 (brown), 1.1 (grey), 1.2 (chartreuse), 1.3 (blue), 1.4 (deep sky blue), 1.5 (cyan), 1.6 (turquoise), 1.7 (purple), 1.8 (pink), 1.9 (magenta), and 2 (navy) $M_{\odot}$ with $Z_{\odot}$. Dashed lines represent stars outside of the theoretical range of application of the \citet{Kopparapu14} prescription.}
  \label{HZevolmassrin}
\end{figure}

Figure \ref{HZevolmassrin} shows the evolution of $R_{\rm{in}}$ and $R_{\rm{out}}$ as a function of time and stellar mass {for all the models at solar metallicity Z = 0.0134}. {As shown in \citet{Porto15}, the evolution of $R_{\rm{in}}$ and $R_{\rm{out}}$ along the PMS depends on the stellar mass. During that phase, these quantities continuously decrease for low-mass stars ($M_* < 0.6 M_{\odot}$), while for higher mass stars they first decrease before increasing when CNO burning reaches equilibrium as the stars approach the ZAMS. Then for all stellar masses, $R_{\rm{in}}$ and $R_{\rm{out}}$ remain approximately constant along the MS before strongly increasing when the stars move towards and along the RGB.}

{The increase of HZ is not linear with the variation of mass i.e. at about 100-200 Myr the difference between $R_{\rm{in}}$ or $R_{\rm{out}}$ of stars separated by 0.1 $M_{\odot}$ strongly increases with mass, and the width of the HZ increases as well for increasing stellar mass. On average, and for a non rotating star with solar metallicity, the width of the HZ is about 0.13 AU for a 0.3 $M_{\odot}$ star and 3.25 AU for a 2 $M_{\odot}$ star.} Variation on stellar mass will also dramatically affect the location of the HZ. At ZAMS, and by increasing the stellar mass from 1 to 2 $M_{\odot}$, the HZL moves outward by more than 400\%. 

\begin{table}
\caption{Size of the HZ as a function of stellar mass.}
\centering
\label{SizeHZ}
\begin{tabular}{ccccc}
\hline
\hline
 ${\Delta}HZ$ & 0.3 $M_{\odot}$ & 1 $M_{\odot}$&  1.4 $M_{\odot}$ & 2 $M_{\odot}$\\
\hline
   ${\Delta}HZ_{\rm{mean}}$ (AU) & 0.13 & 0.86 &  1.68  & 3.25 \\
   ${\Delta}HZ_{\rm{min}}$ (AU) & 0.12  &  0.65 & 1.08 &   1.85 \\
   ${\Delta}HZ_{\rm{max}}$ (AU) & 0.48  &  3.46  & 5.45  &6.63\\
\hline   
\end{tabular}
\end{table}
\begin{figure}[ht]  
\centering
 \includegraphics[angle= -90,width=0.5\textwidth]{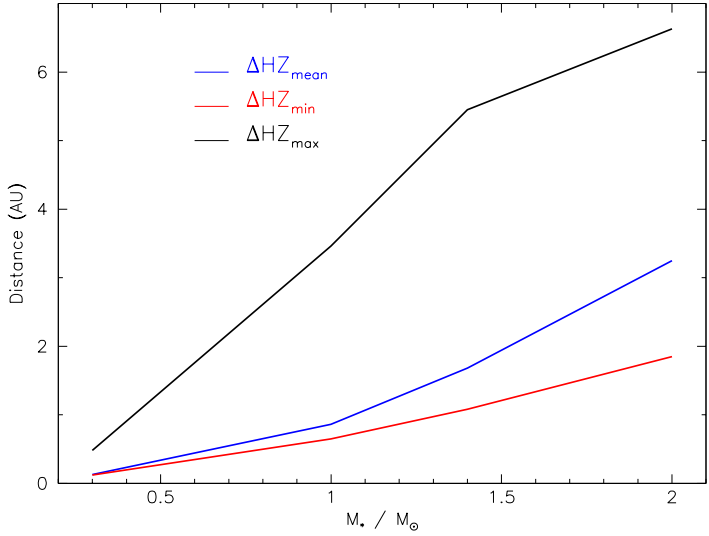}
 \caption{Averaged (blue), minimum (red), and maximum (black), size of the HZ as a function of stellar mass. {These values are compiled over the whole stellar evolution}.}
  \label{SizeHZfig}
\end{figure}
Table \ref{SizeHZ} and Fig. \ref{SizeHZfig} show the {maximum, minimum, and averaged value} of the width of the HZ, where 
\begin{eqnarray}
{\Delta}HZ_{\rm{mean}} = \sum_{i=1}^{n} \frac{R_{\rm{out},i} - R_{\rm{in},i}}{n},
\end{eqnarray}
with $n$ the total number of steps from the stellar model. The minimum is reached at 352, 14, 7.35, and 3.5 Myr, respectively for 0.3, 1.0, 1.4, and 2.0 $M_{\odot}$. {The fact that the width of the HZ of high mass stars is, on average, wider than {that of low mass stars suggests that the probability for a  planet to be found inside the HZ is higher for higher mass stars. However, on the other hand, the duration of the continuously habitable zone strongly decreases for increasing stellar mass (see Sect. \ref{CHZ}) which dramatically decreases the chance for a planet inside the HZ of a high mass star to end up with a complex life.}} The effect of a change of stellar mass on the  stellar temporal features will be somehow the same as a change of metallicity. Increasing mass (decreasing metallicity) will shorten the stellar life.

\section{Continuously habitable zone}
\label{CHZ}

The continuously habitable zone (CHZ) is a region within the HZ where a planet, given its orbital radius, will stay a minimum amount of time. This duration is generally assumed to lie between 2 Gyr \citep{Truitt15} and 4 Gyr \citep{Valle14}. According to \citet{Truitt15}, 2 Gyr is the time required by primordial life form on Earth to significantly alter the atmospheric composition of the planet so that this alteration (i.e. called biosignature) could be detectable if viewed from another planetary system \citep{Kasting93}. On the other hand, 4 Gyr is thought to be the typical timescale for the emergence of complex life such as on Earth.  
{However,} these two timescales are somehow too restraining since they refer to an Earth-like life. Therefore, in the following we do not fix any minimum duration for the CHZ but rather simply look for the HZL in which a potential planet will stay the longest amount of time.
\begin{figure}[ht]
\centering
 \includegraphics[angle= -90,width=0.5\textwidth]{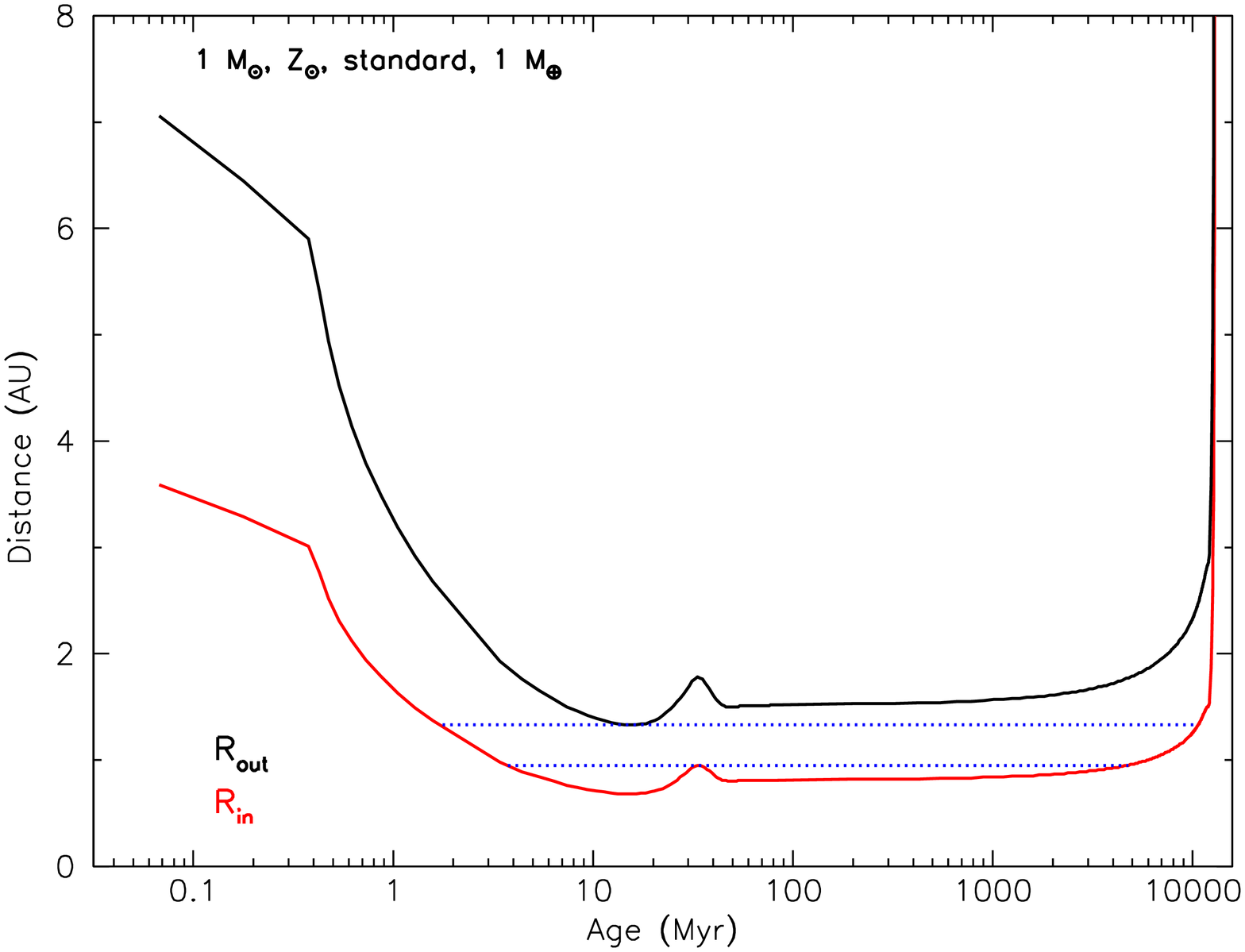} \\
 \includegraphics[angle= -90,width=0.5\textwidth]{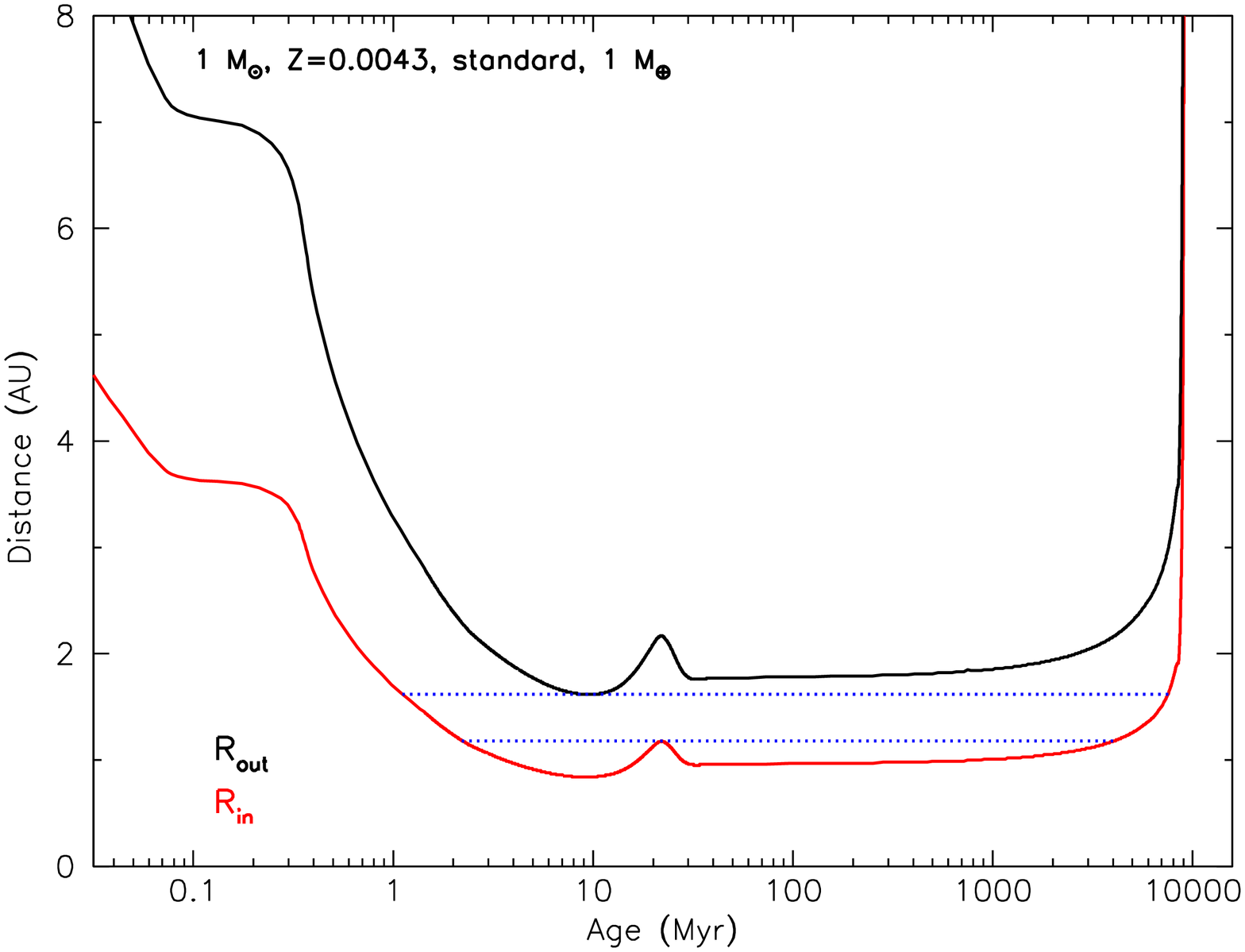}
  \caption{Continuously habitable zone in the case of a 1 $M_{\odot}$ star with solar metallicity and Z=0.0043 (top and bottom panels, respectively). The inner (red) and outer (black) HZ limits were obtained using standard model and the \citet{Kopparapu14} prescription. The blue horizontal lines delimits the area that corresponds to the maximum duration, of a given planet, within the HZ.}
  \label{CHZ1msol}
\end{figure}

\begin{table}[ht]
\caption{Size of the CHZ as a function of stellar mass for solar metallicity. Time (Gyr) = duration of the CHZ near the inner ($\rm{Time_{in}}$) and outer ($\rm{Time_{out}}$) edge of the HZ.}
\label{SizeCHZ}
\begin{center}
\begin{small}
\begin{tabular}{ccccc}
\hline
\hline
Mass & $R_{\rm{in}}$ (AU)  & $\rm{Time_{in}}$ (Gyr) & $R_{\rm{out}}$ (AU)  & $\rm{Time_{out}}$ (Gyr)    \\
\hline
0.3 $M_{\odot}$ & 0.12   & >20 & 0.23  & >20 \\
1 $M_{\odot}$ & 0.95 &   4.77 & 1.33   & 10.67\\
1.4 $M_{\odot}$& 1.96 &  2.15 & 2.27  & 2.73\\
2.0 $M_{\odot}$& 4.27 &  0.99 & 7.0  & 1.09\\
\hline  
\end{tabular}
\end{small} 
\end{center}
\end{table}

{Fig. \ref{CHZ1msol} shows the evolution of the HZ as a function of time for two metallicities ($Z_{\odot}$ and $Z=0.0043$, top and bottom panels respectively). The CHZ is depicted by the dotted blue lines in these figures. It corresponds to the region (expressed in astronomic units) inside the HZ where a planet will potentially stay a maximum amount of time. Since we did not impose in this work any minimum duration for a planet to be inside the CHZ, we also give the potential time that a planet, located either close to the inner or outer edge of the CHZ, will pass inside the CHZ.}

For a solar type star, the CHZ limits (CHZL) will move closer from the star for increasing metallicity (see Fig. \ref{CHZ1msol}). As the stellar lifetime will decrease when decreasing the metallicity (see Sect. \ref{metal}) the CHZL will also last longer for metal rich stars. The time spent within the CHZ is longer for planets close to the outer edge of the HZ and decreases at lower metallicity. However, the width of the CHZL (i.e. the size between the inner and outer edges of the CHZ) increases with decreasing metallicity. Moreover, stars with lower metallicity will enter the CHZ earlier. 

For a comparison purpose, we also put the results for a 5 and 0.1 $M_{\oplus}$ planets around a 1 $M_{\odot}$ and $Z_{\odot}$ star in Table \ref{SizeCHZplanet}. The planetary mass has only a small effect of the location of the CHZL and the duration of the CHZ. However there is a little trend for smaller CHZL and smaller duration of the outer CHZ towards heavier planet.
\begin{table}[ht]
\caption{Size of the CHZ as a function of planetary mass around a solar mass star with solar metallicity. Time (Gyr) = duration of the CHZ }
\label{SizeCHZplanet}
\begin{center}
\begin{small}
\begin{tabular}{ccccc}
\hline
\hline
Planet & $R_{\rm{in}}$ (AU)  & Time (Gyr) & $R_{\rm{out}}$ (AU)  & Time (Gyr)    \\
\hline
0.1 $M_{\oplus}$ & 0.92   & 4.88 & 1.33  & 11.01 \\
1 $M_{\oplus}$ & 0.95 &   4.77 & 1.33   & 10.67\\
5 $M_{\oplus}$& 1.01 &  4.88 & 1.33  & 10.07\\
\hline  
\end{tabular}
\end{small} 
\end{center}
\end{table}

\section{Stellar activity, magnetospheric protection, and habitable zone}
\label{activity}

Stellar and planetary magnetic fields are thought to be generated by {dynamo action within their convective layers} \citep[e.g.][]{Stevenson03,Brun15b}. {On one hand,} in the case of the star, the magnetic field plays a major role {in its overall dynamics and shapes its environment. Magnetic field is also involved in} SDI processes, stellar winds and magnetic braking, and surface activity (starspots, flares, CME's), {and as a consequence has a direct impact on the surrounding planets}. {On the other hand, planetary magnetic fields are} primarily crucial for magnetic protection. For instance, the magnetosphere of the Earth protects us from the radiation and wind emitted by the Sun and transported through its magnetic field. The only star-planet system that is fairly well known is the Sun-Earth's one. This system is not unique and it is though that other 
Star-exoplanet(s) systems should experience such magnetic interactions \citep{Vidotto13,Stu14,Stu15}. {In this context,} constraining both stellar and planetary magnetic activity and magnetic field strength is thus essential when assessing planetary habitability. 

\subsection{Context}

During the last decade, stellar magnetic fields have been quite well documented thanks to 3D topology reconstruction \citep[see e.g. ][]{Donati08,Petit08,Morin10,Mathur15} and {3D dynamo simulations {\citep{Brun04,Browning08,Brown10,Augustson13,Palacios14,Kapyla14,Brun15,Hotta15,Augustson15,Guerrero16}}}. 

More recently the interaction of stellar winds with planetary magnetospheres {have} just started to be intensively studied and modelled \citep[e.g. ][]{Cohen09,Vidotto13,Stu14,Stu15,Vidotto15}. This interaction seems to have a very strong impact on the planetary habitability since a large fraction of the atmospheric component of a planet can be removed by these winds that are thought to be strongly correlated to the X-ray luminosity known to be a good activity proxy \citep{Wood05,Vidotto14,Vidotto16}.

As shown previously, the HZ location and width are not fixed but rather vary together with the stellar parameters along secular evolution. The question is then to know where the HZ is located regarding the stellar activity in order to estimate its impact on habitability.

\subsection{Rossby number and stellar magnetic field generation}

\subsubsection{Generalities}

{The Rossby number quantifies the impact of rotation on the convective motion inside the star through a ratio of the inertial force to the Coriolis force.  Thus, a Rossby number less than one indicates that the Coriolis force is larger than the inertial force.  In agreement with both theory and experiments \citep[e.g.,][]{Durney78,Noyes84,Jouve10,calkins15, cheng15,Brun15b,calkins16}, a low Rossby number typically implies that the convective flows become quasi-two-dimensional and subsequently they generate a stronger and more spatio-temporally coherent kinetic helicity. The kinetic helicity is a critical component for a substantial $\alpha$-effect, which arises due to the action of small-scale vortices twisting the mean toroidal magnetic field into a mean poloidal magnetic field \citep[e.g.,][]{moffat78}.  Moreover, low Rossby number convection usually supports a strong solar-like differential rotation, which in turn gives rise to a strong $\Omega$-effect wherein the differential rotation winds up the mean poloidal magnetic field into a toroidal magnetic field.  Hence, with both the $\alpha$ and $\Omega$ effects likely being more effective at low Rossby number, one could expect more vigorous magnetic field generation within the convection zone and subsequent magnetic activity at the star's surface.}

\subsubsection{Prescriptions}

\begin{figure}[ht]
\centering
 \includegraphics[angle= -90,width=0.5\textwidth]{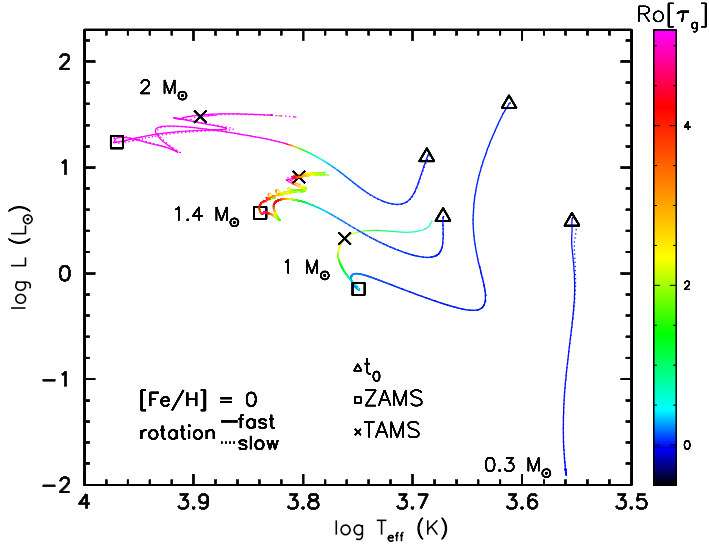}
 \caption{Variation of the global Rossby number (colour-coded) along the evolution tracks in the HRD for the 0.3, 1, 1.4, and 2 $M_{\odot}$, $Z_{\odot}$ models for the fast and slow rotating cases (solid and dashed lines respectively). The triangles, squares and crosses represent the location in the HRD of the age of the first time-step $\rm{t_0}$ in each STAREVOL models, ZAMS and TAMS, respectively.}
  \label{rossbyhp2}
\end{figure}

While stellar rotation has only a small effect on the location of the HZL for the domain in stellar mass we consider (as shown in Fig. \ref{HZrot}), it strongly impacts the stellar activity that is intimately linked to the Rossby regime in which the star is located \citep{Brun05}. As defined before, the Rossby number evaluates the ratio of inertial to Coriolis force. It can be approximated by 
\begin{eqnarray}
R\rm{o} = \frac{P_{\rm{rot}}}{\tau_{\rm{conv}}},
\end{eqnarray}
where $P_{\rm{rot}}$ is the surface rotation period of the star (in days), and $\tau_{\rm{conv}}$ the convective turnover timescale (in days) at a given depth within the stellar convective envelope (when it is present) given by
\begin{eqnarray}
\tau_{\rm{conv}}(r) = \alpha H_p(r)/V_c(r),
\end{eqnarray}
where $\alpha$, $H_p$, and $V_c$ are the mixing-length parameters, local pressure scale height, and the local convective velocity. For a given star, the magnetic activity is expected to increase with decreasing Rossby number. {Figure \ref{rossbyhp2} shows the evolution in the HRD, for three selected stellar masses, of the global Rossby number estimated using rotating models with initial angular velocities as listed in Tab. \ref{vit}.}
The global convective turnover timescale associated to the global Rossby number $R\rm{o}[\tau_g] = P_{\rm{rot}}/\tau_{\rm{g}}$ is defined by 
\begin{eqnarray}
\tau_{\rm{g}} = \int_{R_{\rm{b}}}^{R_{*}} \frac{dr}{V_c(r)},
\end{eqnarray}
with $R_{\rm{b}}$ being the radius at the bottom of the convective envelope and $R_*$ the surface stellar radius.

\subsubsection{Rossby behaviour and correlation with the habitable zone}

{In} Fig. \ref{rossbyhp2}, {we note} three different kinds of behaviour : low-mass star, solar-like star, and intermediate-mass star. While the low-mass star (0.3 $M_{\odot}$) stays in a very high stellar activity regime (low Rossby number) during its whole evolution due to a long convective turnover timescale because of a weak convective flow as the $V_C$ scales as $\sqrt[3]{L_*}$, solar and intermediate-mass stars {will experience a high activity regime on the PMS, that will be followed by a lower activity phase (larger Ro numbers)}. There are two possible reasons for the increase of the Rossby number: either the rotation period $P_{\rm{rot}}$ increases because of magnetic braking by stellar wind during the MS or the convective turnover time $\tau_{\rm{conv}}$ decreases because of the reduction of the size of the convective envelope {or the increase of the amplitude of the convective motions that scale as $\sqrt[3]{L_{*}}$}. These two situations are represented in Fig. \ref{rossbyhp2} by the solar and high-mass stars, respectively.  

In the case of the solar-type stars, the evolution of the Rossby number is mainly controlled by the rotation period. The star maintains a high activity regime during the entire PMS phase and it only starts to decrease after the ZAMS, i.e. when the rotation period starts to increase (smaller rotation rate) because of the magnetic braking \citep[see][and references therein]{GB13,GB15}.

For the intermediate-mass star, the evolution of the Rossby number is controlled by the evolution of the internal structure. Figure \ref{rossbyhp2} shows that the Rossby number already starts to increase during the mid-PMS phase (around $4\times10^5$ years) because of the radiative core development. The Rossby number thus rapidly increases above Ro=1. When the star becomes almost radiative (size of the convective envelope lower than 1\% of the size of the star, around 5 Myr for a 2 $M_{\odot}$)  the Rossby number sharply rises toward large values and hence a likely low activity regime.

\begin{figure}[ht]
\centering
 \includegraphics[angle=-90,width=0.5\textwidth]{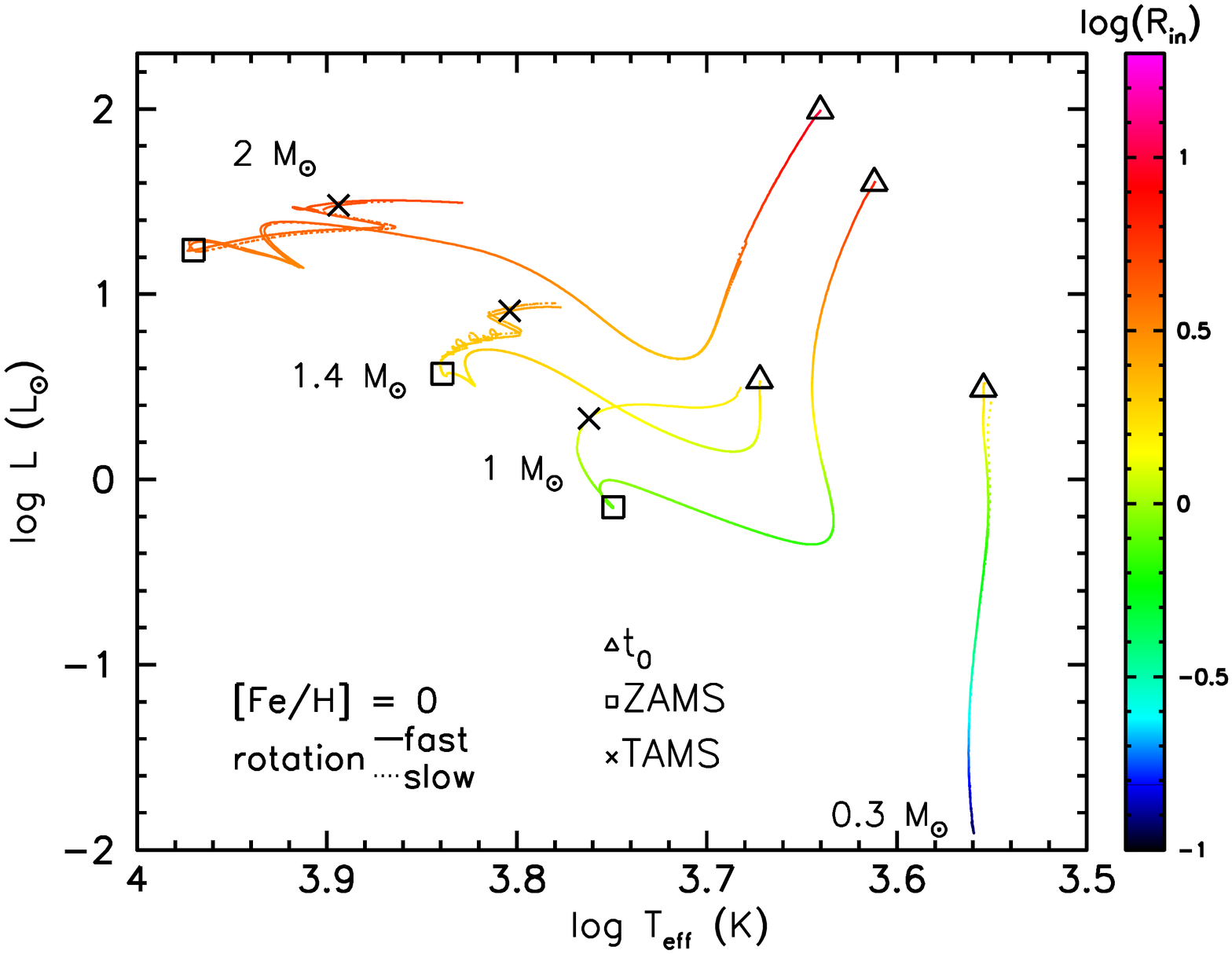} \\
  \includegraphics[angle=-90,width=0.5\textwidth]{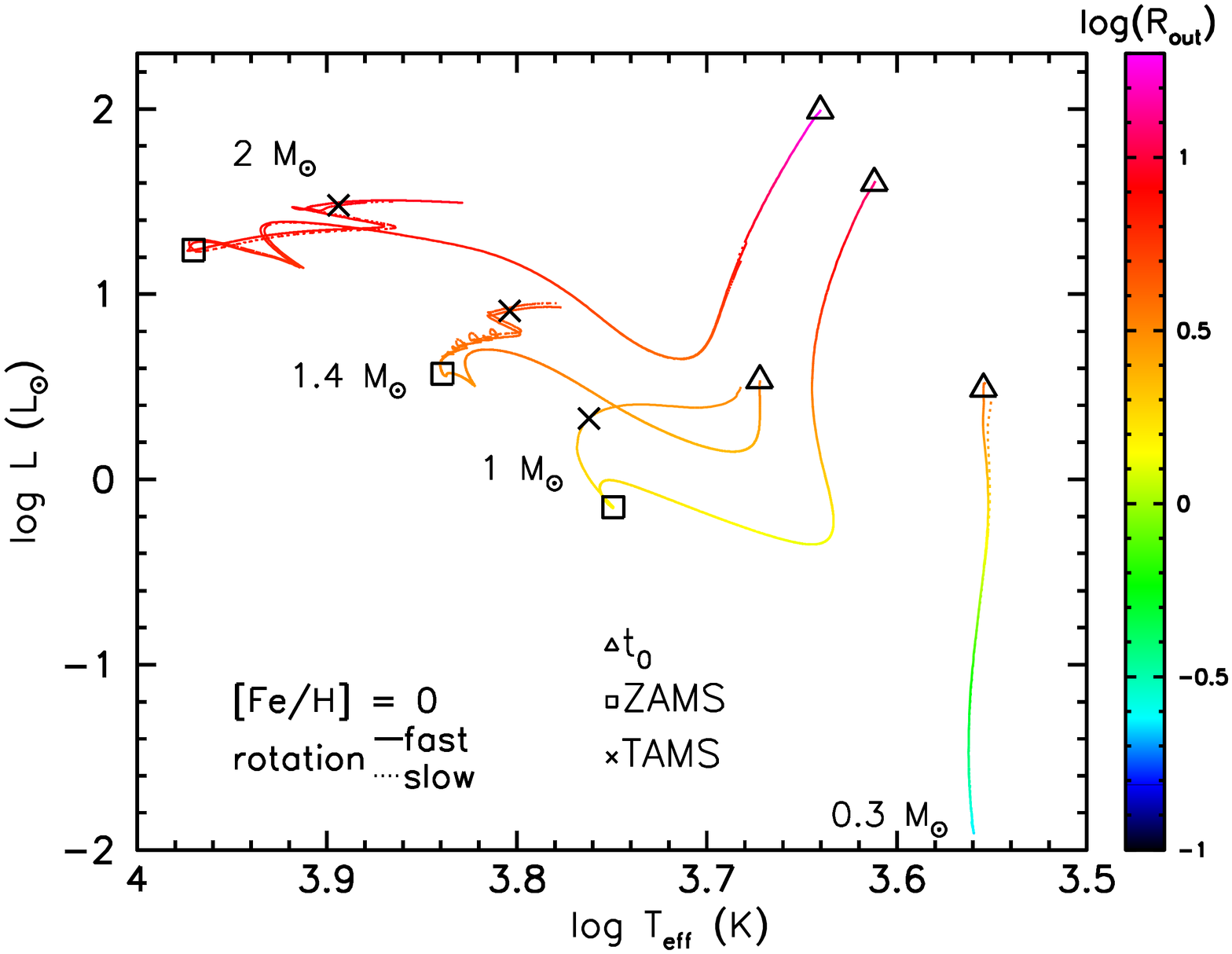}
  \caption{HRD of a {0.3}, 1, 1.4, and 2 $M_{\odot}$ stars with solar metallicity in fast (solid line) and slow (dashed line) rotation regime. The coloured gradient depicts the variation of $R_{\rm{in}}$ and $R_{\rm{out}}$ (top and bottom, respectively), expressed in AU, along the stellar evolution. The triangles, squares and crosses represent the location in the HRD of the age of the first time-step $\rm{t_0}$ in each STAREVOL models, ZAMS and TAMS, respectively.}
  \label{HZincolor}
\end{figure}

Figure \ref{HZincolor} depicts the evolution of the inner and outer limits of the HZ, respectively,  along the tracks in the HRD of a 0.3, 1, 1.4, and 2 $M_{\odot}$ stars with solar metallicity in fast and slow rotation regimes from the PMS to the end of the MS phase.  $R_{\rm{in}}$ and $R_{\rm{out}}$ decrease during the Hayashi track (i.e. the HZ become closer to the star regardless of the stellar mass) while the {global} Rossby number is quite small and around $ Ro \approx 10^{-3}-10^{-4}$ (see Fig. \ref{rossbyhp2}). 

As the 1, 1.4, and 2 $M_{\odot}$ stars evolve toward the ZAMS, their luminosity and effective temperature slowly increase which produces a small rise of both $R_{\rm{in}}$ and $R_{\rm{out}}$. Then, on MS phase, the HZL follow the intrinsic evolution of stellar luminosity and slightly increase all the way up to the end of the MS. Stars above 0.5-0.6 $M_{\odot}$ will reach a state of small HZL during the mid-PMS phase. 

For the 0.3 $M_{\odot}$ star, this evolution is quite different. Since the stellar luminosity continuously decreases at almost constant effective temperature from the PMS to the end of the MS. 

The evolution of the stellar activity, that is estimated here simply thanks to the Rossby regime in which the star is, strongly depends on the stellar mass and more specifically on the internal structure of the considered star. 

{For the 1, 1.4, and 2 $M_{\odot}$ stars, the stellar activity is high (i.e. Rossby number $<$ 1) along the PMS phase during which the HZL are the closest from the star. For these stars, by opposition, the stellar activity is relatively low during the entire MS (and late PMS phase for the 2 $M_{\odot}$ star) phase and may then not impact the long term development of life during the earliest stages of the continuously habitable zone (hereafter CHZ, see Sect. \ref{CHZ}).} 

{Conversely the 0.3 $M_{\odot}$ star stays in a high activity regime during its whole life while the HZ continuously moves closer from the star.}
 
While there is no definite correlation between the activity regime in which the star is and the location of the HZL, the stellar activity must be taken into account, and more specifically for M dwarfs, to assess the habitability of a planet since {currently} almost all detected potentially habitable planet orbit low-mass stars (i.e. around 0.3 $M_{\odot}$). For this stellar mass regime, and as we showed above, the stellar activity cannot be neglected and may have strong impacts on the life development/complexification.

{Moreover, short term extreme events, such as Carrington-like and superflares events \citep{Maehara12}, can occur. Thus using only the Rossby number to assess space weather condition is a zero order approach that will need to be further refined is future studies.}

\subsection{Magnetic flux and magnetospheric protection}

{In addition to be inside the HZ, a planet also needs to possess an effective magnetospheric protection to be considered as habitable. There are few questions that are worth to answer : what is the maximum stellar magnetic field (thus magnetic flux) that a planet found inside the HZ can handle given its own magnetic field? What is the minimum planetary magnetic field required to handle a sun-like star magnetic field evolution? } 

{General theoretical robust scaling laws for the amplitude of magnetic field as a function of parameters computed only thanks to stellar evolution codes are not yet available in the literature.} {Either the magnetic field {amplitude is computed} from complex 3D numerical simulations {and simplified asymptotic balances} \citep[e.g.][]{Brun15,Augustson15} or from simplified relation by assuming the magnetic field at the thermal equilibrium with the stellar photosphere \citep[e.g.][]{KV00,Cranmer11}. Therefore, we rather directly use the observed trends of the magnetic field strength as a function of time, rotation and Rossby number.}

{\citet{Folsom16} provide a compilation of magnetic field measurements from three large spectropolarimetric surveys: HMS (the History of the Magnetic Sun in the framework of the Toupies\footnote{Toupies: TOwards Understanding the sPin Evolution of Stars.} project), BCool, and  MaPP\footnote{MaPP: MAgnetic Protostars and Planets.}. Their figure 9 displays clear trends of the mean large-scale magnetic field strength with age, Rossby, and also rotation (but only for high rotation period). }

{In the framework of an exoplanet orbiting a star with a large enough orbital radius (larger than few tenth of stellar radius) the magnetic field to take into account is the large scale magnetic field. \citet{Folsom16} found that 
\begin{eqnarray}
B_* = 1680.9 \times t^{-0.74308} \rm{~G},
\end{eqnarray}
where $t$ is the age expressed in Myr, and
\begin{eqnarray}
\label{bobs}
B_* \propto R\rm{o}^{-1}.
\end{eqnarray}
We normalized this latter expression so as to reproduce the Sun's magnetic field strength:
\begin{eqnarray}
B_* = B_{\odot} \frac{R\rm{o}_{\odot}}{Ro_*},
\end{eqnarray}
with $B_{\odot}=$ 2 Gauss, $R\rm{o}_{\odot}=$ 1.96, {while} $R\rm{o}_*$ is estimated using $\tau_{g}$. }

{\citet{Vidotto13} provided a direct relation between the minimum planetary magnetic field, required for an effective magnetic protection, and the stellar magnetic field :
\begin{eqnarray}
B^{\rm{min}}_p \simeq 16052 P_{\rm{B},*}(R_{\rm{orb}})^{1/2},
\label{bmin}
\end{eqnarray}
where
\begin{eqnarray}
P_{B,*}(R_{\rm{orb}}) = \frac{B_{\rm{SS}}^2}{8\pi} \left(   \frac{R_{\rm{SS}}}{R_{\rm{orb}}} \right)^4,
\label{pb}
\end{eqnarray}
is the magnetic pressure with $R_{\rm{SS}}$ the source surface radius \citep[assumed to be 2.5 $R_*$, which is known to not be valid for all ages and rotation rates, see][]{Reville15b}, $B_{\rm{SS}}$ the magnetic field at $R_{\rm{SS}}$, $R_{\rm{orb}}$ the orbital radius of the exoplanet, and $B^{\rm{min}}_{\rm{p}}$ the minimum planetary magnetic field required for an effective magnetic protection.} Here we assume a simple region of interaction and we neglect the planetary magnetic field orientation \citep{Saur13,Stu15} as well as the activity cycle and grand maxima phase.

{We considered an Earth mass planet situated at 1 AU. By coupling Eqs. (\ref{bmin}) and (\ref{pb}), the minimum planetary magnetic field can be expressed as a function of stellar and planetary parameters as follows
\begin{eqnarray}
B^{\rm{min}}_p \simeq 16052  \left[ \frac{B_{*}^2}{8\pi} \left(   \frac{2.5R_{*}}{1 AU} \right)^4\right]^{1/2}.
\end{eqnarray}
Note that $B^{\rm{min}}_{\rm{p}}$ is proportional to the squared stellar radius.}

\begin{figure}[ht]
\centering
 \includegraphics[angle=-90,width=0.45\textwidth]{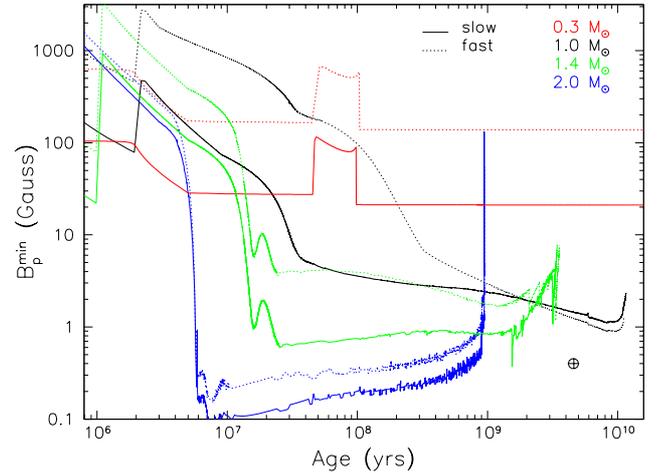}
  \caption{Minimum planetary magnetic field required by a 1 $M_{\oplus}$ orbiting a 0.3, 1, 1.4, and 2 $M_{\odot}$ at 1 AU to produce an effective magnetic protection. Solid and dashed lines represent slow and fast rotators, respectively. The Earth is represented by the $\oplus$ symbol.}
  \label{Magflux}
\end{figure}

{At the age of the Sun, an Earth-like planet needs a magnetic field of at least 0.8 G to be protected from a solar-like magnetic field of 1.88 G (see Fig. \ref{Magflux}), which is consistent with the measured values of the Earth ($B_{\oplus} \approx 0.5$ G) and Sun ($B_{\odot} \approx 2$ G) magnetism. {Note that these small differences between numerical and observed magnetic fields are due to the employed observed magnetic prescription that only provide order of magnitude magnetic field.} Because more massive stars are less active on the MS (see Fig. \ref{rossbyhp2}), $B^{\rm{min}}_{\rm{p}}$ will also globally decrease towards higher masses. However, since $B^{\rm{min}}_{\rm{p}}$ is proportional to the squared stellar radius, the minimum planetary magnetic field will increase during the main sequence as the stellar radius increases.
While the evolution of $B^{\rm{min}}_{\rm{p}}$ in the case of the 1 $M_{\odot}$ star is dominated by the evolution of the stellar magnetic field controlled by the rotation period of the star through the Rossby number, in the case of the 2 $M_{\odot}$ star this evolution is controlled by the evolution of the stellar radius that slightly increases during the MS phase while the magnetic field remains quite constant.}

The minimum planetary magnetic field for the 1 $M_{\odot}$ first starts to decrease up to 2 Myr when the radiative core appears inside the star. This stable core development decreases the convective turnover timescale and thus decreases the Rossby number of the star, which leads to the increase of the stellar magnetic field strength as well as $B^{\rm{min}}_{\rm{p}}$. This diminution then continues up to the ZAMS where the size of the core will remain almost constant. After the ZAMS the decrease of $B_*$ (and $B^{\rm{min}}_{\rm{p}}$) is entirely controlled by the rotation, as $R_*$ is almost fixed, that continuously decreases because of the stellar magnetic wind braking. Beyond the MS turn-off and during the RGB phase both magnetic fields slightly increase as the radius of the star increases so as to maintain a surface hydrostatic equilibrium. 

The evolution of $B^{\rm{min}}_{\rm{p}}$ for the 2 $M_{\odot}$ follows more or less the same trends except that the internal structure of this star is fixed at the beginning of its evolution. $B^{\rm{min}}_{\rm{p}}$ first starts to decrease because of the stellar contraction, then reaches a plateau at ZAMS that will last until the red-giant phase where $B^{\rm{min}}_{\rm{p}}$ will sharply increase. 

In the case of the 0.3 $M_{\odot}$ star a radiative core starts to develop just before 100 Myr and will last for only 50 Myr. The growth of the core sharply decreases the convective turnover time (as the size of the convective envelope decreased) producing the rise of the stellar and thus planetary magnetic field in our simplified approach. As a consequence, the minimum planetary magnetic field rises up to 100 G during this brief phase. As a 0.3 $M_{\odot}$ is fully convective during almost its entire life and as the magnetic braking is not efficient during the MS, the minimum planetary magnetic field remains quite constant throughout the entire evolution of such low-mass star.

During the MS, the minimum magnetic field $B^{\rm{min}}_{\rm{p}}$ in the case of a solar-type star is about one order of magnitude lower than $B^{\rm{min}}_{\rm{p}}$ for a M-dwarf star.

{Inversely, if we fix the planetary magnetic field we can study the evolution of the maximum stellar magnetic field that a planet can handle:}
\begin{eqnarray}
\label{bmax}
B^{\rm{max}}_* \simeq    \frac{B^{\rm{min}}_{\rm{p}}}{16052}   \left[ 8\pi \left(   \frac{1 AU}{2.5R_{*}} \right)^4\right]^{1/2}.
\end{eqnarray}

\begin{figure}[ht]
\centering
 \includegraphics[angle=-90,width=0.45\textwidth]{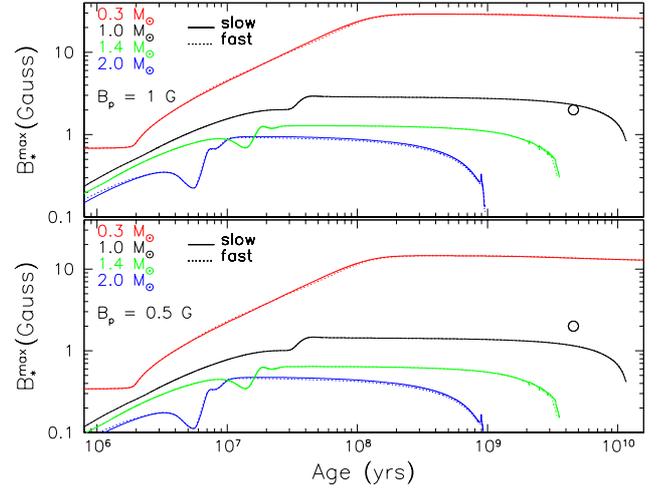}
  \caption{Maximum stellar magnetic field that a 1 $M_{\oplus}$ orbiting a 0.3, 1, 1.4, and 2 $M_{\odot}$ star at 1 AU can sustain given its fixed magnetic field. Solid and dashed lines represent slow and fast rotators, respectively. The Sun is represented by the $\odot$ symbol.}
  \label{Magstar}
\end{figure}

{Figure \ref{Magstar} shows the temporal evolution of $B^{\rm{max}}_*$ as a function of the stellar mass for two planetary magnetic field strengths (0.5 and 1 Gauss).}
{As the star evolves, its radius first decreases which is translated, at fixed planetary magnetic field, by an increase of the maximum stellar magnetic field that the planet can support (since $B^{\rm{max}}_*$ evolves as $R_*^{-2}$, see Eq. \ref{bmax}). During the MS this maximum magnetic field reaches a plateau as the stellar internal structure remains almost constant. During the RGB, and following the stellar expansion, $B^{\rm{max}}_*$ slowly decreases. As expected a more magnetized planet can handle a more intense stellar magnetic field strength.} Note here that Eq. (\ref{bmax}) only takes into account of the structural aspect (through the evolution of $\rm{R_*}$) and no dynamical evolution. Moreover, the effect of the orientation of the planetary magnetic field with respect to that of the star is here also not taken into account.

\section{Conclusions}
\label{conc}

To assess the habitability of a class I exoplanet (i.e. with surface liquid water), the HZL first needs to be precisely defined. {In the literature the HZ is considered at a specific time of the evolution (i.e., considering the properties of the present-day Sun). However, we show in this paper that the variations of the HZL are important as stars evolve, and that the HZL depend on stellar mass and metallicity because they} strongly affect the stellar luminosity and effective temperature. These two quantities are marginally affected by a change in rotation rate.  {However, stellar rotation will have a major role on tidal and magnetic torques evolution \citep{Mathis15,Stu15,BM16} that will then dramatically affect the orbital evolution of a given planet.}
We have shown in this work that a correlation exist between the  stellar activity and location of the HZ along the stellar evolution, indeed the HZL are the closest from the star when this latter is in a high active regime. Eventhough the impact of stellar activity on the HZL is not straightforward to assess, it may impact the planet's atmospheric conditions which in turn will impact the location of the HZL. In this work we showed that M-dwarf stars possess a strong stellar activity during their entire life, while more massive stars (solar-type and intermediate mass star) only exhibit strong activity during their early evolution (i.e. during their PMS phase). As the stellar mass decreases its magnetic field strength globally increases, at least during the MS phase during which a star-planet system will pass almost its entire evolution. As a consequence, during the MS the minimum magnetic field required by an exoplanet for an efficient magnetospheric protection will increase for decreasing mass. This could pose a problem in finding habitable planets as exoplanets inside the habitable zone are currently almost only found around M-dwarf stars. 

With this work we provide better constraints on the HZL by using the stellar evolution code STAREVOL that incorporates up-to-date mechanisms of stellar evolution physics. The final purpose of this study is to provide the community with an on-line tool dedicated to the study of the evolution of the habitability (link to the HZ and planet's orbital motion) of a potential exoplanet. {This tool will then be very useful for example for the CHEOPS \citep{CHEOPS}, TESS \citep{TESS}, PLATO \citep{Plato}, and SPIRou \citep{Moutou15} community}. Thanks to the models presented in this paper we can predict the HZL for numerous star-planet configurations (and for three planetary masses), and can already provide the community with a dedicated grid of stellar models to estimate this quantity. 

Additionally, the evolution of the planet's orbital motion is expected to strongly impact the habitability by moving the planet inside or outside the HZ. Detailed analysis of tidal dissipation in both planet and star is then crucial to study this evolution \citep[e.g.][]{MR13,Ogilvie14}. This mechanism will be soon implemented in the STAREVOL code using the formalism of \citep{Ogilvie13} and \citet{Mathis15}, and the corresponding results will be presented in a forthcoming paper. 

\begin{acknowledgements}
We thank the referee for her/his comments which allow us to improve the article. The authors are grateful to K. Augustson for valuable discussions and inputs on stellar activity. This work results within the collaboration of the COST Action TD 1308 and by the grant ANR 2011 Blanc SIMI5-6 020 01 ``Toupies: Towards understanding the spin evolution of stars'' (\url{http://ipag.osug.fr/Anr_Toupies/}). The authors acknowledge financial support from the Swiss National Science Foundation (FNS) and from the French Programme National de Physique Stellaire PNPS of CNRS/INSU. A.S.B. acknowledges support by CNES through Solar Orbiter and Plato fundings as well as ERC STARS2 207430, Solar Predict 640997 and FP7 Spaceinn grants. S. Mathis acknowledges funding by the European Research Council through ERC SPIRE grant 647383.
\end{acknowledgements}


\bibliographystyle{aa}
\bibliography{habv6arxiv}

\end{document}